\RequirePackage{rotating}
\documentclass [11pt]{article}

\usepackage{rotating}
\usepackage{setspace}
\usepackage{graphicx}
\usepackage{amssymb}
\usepackage{mathtools}
\usepackage{multirow}
\usepackage{pdflscape}
\usepackage{array}
\usepackage{endfloat}
\usepackage{natbib}
\RequirePackage{fix-cm}
\usepackage{geometry}
\newcommand{\bA}{{\bf A}}
\newcommand{\bD}{{\bf D}}
\newcommand{\bM}{{\bf M}}
\newcommand{\bZ}{{\bf Z}}
\newcommand{\bK}{{\bf K}}
\newcommand{\bk}{{\bf k}}
\newcommand{\bh}{{\bf h}}
\newcommand{\bx}{{\bf x}}
\newcommand{\bzero}{{\bf 0}}

\newcommand{\blambda}{\pmb{\lambda}}
\newcommand{\bomega}{\pmb{\omega}}

\newtheorem{Proposition}{Proposition}
\newtheorem{Lemma}{Lemma}

\newcommand{\finpreuve}{\hfill $\Box$}

\begin{document}

\title{Half-tapering strategy for conditional simulation with large datasets}
\date{}
\author{D. Marcotte$^a$  and D. Allard$^b$\\
\small{$^a$D\'{e}partement des g\'{e}nies civil, g\'{e}ologique et des mines, Polytechnique Montr\'{e}al,}\\ 
\small{C.P. 6079 Succ. Centre-ville, Montr\'{e}al, (Qc), Canada H3C 3A7}\\ 
\small{Tel: +1 (514) 340-4711 ext. 4620}\\
\small{Fax: +1 (514) 340-3970}\\
\small{denis.marcotte@polymtl.ca}\\
\small{$^b$Biostatistique et Processus Spatiaux (BioSP), INRA,}\\
\small{84914 Avignon, France}\\
\small{Tel: +33 (0) 432 72 21 71}\\
\small{Fax: +33 (0) 432 72 21 82}\\
\small{denis.allard@inra.fr}
}

\maketitle

\begin{abstract}Gaussian conditional realizations are routinely used for risk assessment and planning in a variety of Earth sciences applications. Assuming a Gaussian random field, conditional realizations can be obtained by first creating unconditional realizations that are then post-conditioned by kriging. Many  efficient algorithms are available for the first step, so the bottleneck resides in the second step. Instead of doing the conditional simulations with the desired covariance (F approach) or with a tapered covariance (T approach), we propose to use the taper covariance only in the conditioning step (Half-Taper or HT approach). This enables to speed up the computations and to reduce memory requirements for the conditioning step but also to keep the right short scale variations in the realizations. A criterion based on mean square error of the simulation is derived to help anticipate the similarity of HT to F. Moreover, an index is used to predict the sparsity of the kriging matrix for the conditioning step. Some guides for the choice of the taper function are discussed. The distributions of a series of 1D, 2D and 3D scalar response functions are compared for F, T and HT approaches. The distributions obtained indicate a much better similarity to F with HT than with T.      
\end{abstract}

\medskip


\newpage

\section{Introduction}
Conditional simulations are  routinely used in Earth sciences applications to assert the distribution of responses that are non-linearly related to the state variables of the system under study. For the specific case of Gaussian random fields, the general approach is to produce many different realizations of the system, each being consistent with the chosen covariance function and the observed data points, and then to apply a physical forward model to get the response variables corresponding to the simulated system state. Conditioning to observed data is usually done by post-conditioning by kriging although some algorithms like sequential Gaussian simulation (SGS) can perform the conditioning directly as the simulation progresses. However, SGS has to use local neighborhoods, hence introducing discontinuities and only approximating the desired covariance \citep{emery2004,emery2008,emery2015,safikhani2016assessing}.  

Given the availability of increasingly large datasets, conditional simulation algorithms still face the size problem. When tens of thousands conditioning data are available, it is hardly possible to produce conditional simulations without some form of approximation.  At least five different approaches were proposed to alleviate the size problem: a) the classical approach of conditioning locally, see e.g. \cite{chiles2012}, b) the fixed rank kriging \citep{cressie2008}, c) tapering of the covariance functions \citep{furrer2006}, d) using a combination of b) and c) \citep{gneuss2013,sang2012} and e) using covariance functions that possess sparse precision matrix (inverse of the covariance matrix) when defined on a torus \citep{lindgren2011}. Approaches b), c) and d) were proposed in the context of prediction rather than simulation, but they can obviously be used for the post-conditioning by kriging. 

All the above approaches have their caveats. With a), discontinuities are introduced as the local neighborhood changes. Moreover, the simulated covariance is distorted compared to the desired covariance. Using b) amounts to keep only the low frequency components of the covariance function which can produce overly smoothed realizations. Approach c) is dependent on the range of the taper function compared to the effective range of the covariance function. When the range of the taper function is larger than the effective range, covariance distortion is reduced but so is the sparsity of the tapered covariance matrix. On the contrary, when the range of the taper function is chosen small, the resulting sparsity is important but so is the distortion of the covariance function. Approach d) appears more complex as one has to split the prediction step in two parts corresponding to a long range-short range decomposition. The long range component is  estimated by fixed-rank kriging and the short range component is estimated by tapering the covariance of the residuals. Finally, although being extremely efficient in some cases, approach e) has limitations, in particular because the Markov property on which it relies has been firmly  established only for Mat\'ern covariances.

Many efficient algorithms exist to produce unconditional Gaussian fields. As a well known example, the turning bands method (TBM) \citep{matheron1971,lantuejoul_geostatistical_2002,emery2006,emery2015spectral} has the lowest computational complexity with $O(N)$, where $N$ is the number of simulated points. Moreover, TBM has no practical memory limitations as points to simulate can be simulated by subsets of any desired size. Alternative methods include spectral methods \citep{shinozuka1972,emery2015spectral} and their more recent FFT implementation on grids, using an approach based on circulant embedding of block Toeplitz covariance matrices \citep{chan1999simulation}. We do not detail any further non conditional simulation methods.

Hence, the bottleneck for large conditional simulation lies in the post-conditioning part. This contribution examines in details a suggestion made by \cite{stein2013}: produce unconditional simulations with efficient algorithm suitable for the desired  covariance and use the tapered covariance only for the post-conditioning part. We call this approach half-tapering (HT). We expect this approach to better reproduce the short scale variations than do conditional simulations using the tapered  covariance (T) for both parts. Consequently, the forward model uncertainty should be better approximated with the HT approach than with the T approach.  Although, for simplicity, only stationary tapers are considered, our theoretical results (propositions 1 to 3) apply as well to non-stationary and/or non-isotropic tapers constructed following \cite{bolin2016}.

We emphasize that we assume that the covariance function is known, i.e. it has been estimated or chosen by the user. Therefore, we do not consider at all the problem of using a tapering approach for the estimation of the covariance function.
In our conditional simulation setting, the user expects an accurate reproduction of the output statistics for the known covariance.
There is a plethora of possibilities for those outputs, and we will consider typical examples in 1D, 2D and 3D. 

After reviewing some results on the tapered covariance approach for prediction in Section 2, we adapt and extend to the context of conditional simulation the measure of similarity between tapered vs full covariance (F) predictions proposed by \cite{stein1993,stein1999} and \cite{furrer2006}. Then, in Section 3, we derive formulas to predict sparsity of the covariance matrix based on the tapered covariance. In Section 4, different non-linear forward models applied on 1D, 2D and 3D simulations are used to assess the similarity of HT and T response distributions with those obtained using the desired covariance (F).
We finish with some elements of discussion in Section 5.

\section{Methodology}

\subsection{Set-up and notations}

Let us consider a zero mean Gaussian random field $Z(\bx),\ \bx \in D \subset \Re^d,\ d \geq 1$ with stationary covariance function
$$C_0(\bh)  = \hbox{Cov}\{Z(\bx),Z(\bx+\bh) \}, \quad h \in \Re^d.$$
Assuming $C_0$ is known and given sample values  $\bZ_n = (Z(\bx_1),\dots,Z(\bx_n))'$, the Best Linear Unbiased Predictor at a point $\bx \in D$, also referred to as kriging in the geostatistics literature, is:
$$Z^*_{C_0}(\bx) = \bZ_n' \blambda_{C_0}(\bx), \quad \hbox{with} \quad \blambda_{C_0}(\bx) = \bK_0^{-1}\bk_0(\bx),$$
where $\bK_0$ is the $n \times n$ matrix with element 
$[\bK_0]_{ij} = C(\bx_i-\bx_j)$ for $1 \leq i,j \leq n$, $\bk_0(\bx)$ is the vector of covariances 
$(C(\bx-\bx_1),\dots,C(\bx-\bx_n))'$ and $\bA'$ denotes the transpose of  matrix $\bA$.

We will further denote $C_{T,\theta}( \bh),\ \bh \in \Re^d$ a tapering function, i.e. a correlation function with compact support. The taper is identically equal to zero outside a particular range $\theta$, typically the ball of radius $\theta$ in $\Re^d$. Methods to construct compactly supported positive definite functions with chosen order of differentiability have been proposed as early as \cite{bohman1960} and \cite{matheron1965}. Other constructions were proposed later in a different context \citep{wendland1995,wu1995}.  \cite{gneiting2002} recognized that Wu's construction is equivalent to Matheron's \it{clavier sph\'erique} \normalfont {for} which specific members are the spherical, cubic and penta covariance models, see Table \ref{tab:covar}. Other tapers are provided by the Wendland functions  \citep{wendland1995} and the Bohman construction \citep{bohman1960}. 

The tapered covariance function is then the product of the original covariance function by the taper
$$ C_1(\bh) = C_0(\bh) C_{T,\theta}(\bh),\ \bh \in \Re^d.$$
The resulting tapered covariance matrix, $K_1$, defined by $C_1$ is thus the Hadamard product matrix
$\bK_1 = \bK_0 \odot \bK_T$, where $[\bK_1]_{ij} = [\bK_0]_{ij}. [\bK_T]_{ij}$ and $\bK_T$ is the matrix with elements $[\bK_T]_{ij} = C_{T,\theta}(\bx_i-\bx_j)$ for $1 \leq i,j \leq n$. The tapered covariance function is definite positive, since the product of positive definite functions is positive definite by Schur's product theorem. By construction, the matrix $\bK_1$ has a high proportion of zero elements when $\theta$ is small compared to pairwise distances between data locations. The taper function must be tailored to the particular desired covariance function, in particular in terms of range and regularity at the origin. We will return to this issue later. 

\subsection{MSE criterion}
 The MSE at point $\bx \in D$, also known in geostatistics \citep{chiles2012} as the kriging variance, is 
\begin{equation}
MSE(\bx,C_0) =\sigma^2 _{k,C_0}(\bx)  = \hbox{Var}_0\{Z_{C_0}^*(\bx)-Z(\bx)\} =\sigma_0^2-\bk_0(\bx)' \bK_0^{-1} \bk_0(\bx)
\label{equ:tail1}
\end{equation}
where $\sigma^2_{k,C_0}(\bx)$ denotes the simple kriging variance obtained with $C_0$ and $\sigma^2_0$ is the variance of $Z(\cdot)$. In Eq. (\ref{equ:tail1}), the variance is computed for the underlying model $C_0$, hence the notation $\hbox{Var}_0$. When the prediction is done with a plug-in covariance function $C_1$ whereas the true covariance is $C_0$, the kriging weights are obtained by solving:
\begin{equation}
\blambda_{C_1}(\bx)=\bK_1^{-1} \bk_1(\bx)
\label{equ:tail2a}
\end{equation}
where $\bk_1(\bx)$ and $\bK_1$ are respectively the right member and the kriging matrix computed with $C_1$. Hence, one obtains:
\begin{eqnarray}
MSE(\bx,C_1) = \hbox{Var}_0\{Z^*_{C_1}(\bx) - Z(\bx)\}&=& \sigma_0^2-2\blambda_{C_1}(\bx)' \bk_0(\bx)+\blambda_{C_1}(\bx)' \bK_0 \lambda_{C_1}(\bx) \nonumber\\
&= &\sigma_0^2-2 \bk_1(\bx)' \bK_1^{-1} \bk_0(\bx) \\
& & + \bk_1(\bx)' \bK_1^{-1} \bK_0 \bK_1^{-1} \bk_1(\bx).
\label{equ:tail2}
\end{eqnarray}
Since  $MSE(\bx,C_0)$ is the kriging variance, i.e. the variance of the Best Linear Unbiased Predictor, one has necessarily $MSE(\bx,C_1) \geq MSE(\bx,C_0)$, which we state and prove formally below.

\begin{Proposition}
\label{prop:prop1}
In the setting above, one has $MSE(\bx,C_1) \geq MSE(\bx,C_0)$ for all $\bx \in D$.
\end{Proposition}

\noindent{\bf Proof} From the definitions of $MSE(\bx,C_1)$ and $MSE(\bx,C_0)$ one has, 
\begin{eqnarray}
\Delta_{MSE}(\bx,C_0,C_1) &= &MSE(\bx,C_1) - MSE(\bx,C_0)\\
 & = & \bk_1(\bx)' \bK_1^{-1} \bK_0 \bK_1^{-1} \bk_1(\bx) 
-2\bk_1(\bx)' \bK_1^{-1} \bk_0(\bx) + \bk_0(\bx)' \bK_0^{-1} \bk_0(\bx) \nonumber\\
& = & (\blambda_{C_1}(\bx) - \blambda_{C_0}(\bx))' \bK_0 (\blambda_{C_1}(\bx) - \blambda_{C_0}(\bx)) \geq 0.
\label{equ:Delta_MSE}
\end{eqnarray}
The last expression is always nonnegative since $\bK_0$ is a covariance matrix, hence positive semi-definite.  \finpreuve

It is worth recalling that both MSE can be computed without actually knowing data values. Only the point locations and the covariance function are needed.  The two MSEs will be different unless $\blambda_{C_1}(\bx) = \blambda_{C_0}(\bx)$. 
The (theoretical) mean square error of prediction (MSE) has been used  in \cite{furrer2006} based on \cite{stein1993} and \cite{stein1999} as a measure of discrepancy between the original (supposed true) covariance function $C_0$ and the covariance function $C_1$ used to perform the prediction.

\subsection{MSE criterion for simulation}
The MSE criterion for prediction of Eq. (\ref{equ:tail1}) and (\ref{equ:tail2}) can be adapted for simulation. For this, it is useful to adopt the decomposition corresponding to post-conditioning by kriging:
\begin{equation}
Z_{cs}(\bx)=Z^*(\bx)+(Z_s(\bx)-Z_s^*(\bx))
\label{equ:post}
\end{equation}
where $Z_{cs}$ is the conditional simulation, $Z_s$ is the unconditional simulation independent on $Z(\cdot)$, $Z^*$ is the kriging with data, and $Z_s^*$ is the kriging with simulated values at data points. The prediction error of the conditional simulation is then written:
\begin{equation}
e(\bx)=Z(\bx)-Z_{cs}(\bx)=(Z(\bx)-Z^*(\bx))-(Z_s(\bx)-Z_s^*(\bx)).
\label{equ:erreur}
\end{equation}
The right hand expression combines two independent kriging errors, one for the field itself, the other one for the simulated field. Hence when both errors are obtained using covariance $C_0$, one obtains the classical result \citep{chiles2012}
\begin{equation}
MSE^s(\bx,C_0)=2 \sigma^2_{k,C_0}(\bx).
\label{equ:MSEA}
\end{equation}
When using a conditional simulation entirely computed with the covariance $C_1$, whereas the true covariance function is $C_0$, one has:
\begin{equation}
MSE^s(\bx,C_1)=MSE(\bx,C_1)+\sigma^2_{k,C_1}(\bx)
\label{equ:MSEB}
\end{equation}
where $\sigma^2_{k,C_1}$ is the kriging variance computed with covariance $C_1$. 
When unconditionally simulating with covariance $C_0$ and then post-conditioning with covariance $C_1$, one obtains:
\begin{eqnarray}
MSE^s(\bx,C_0,C_1) & = & 2MSE(\bx,C_1) 
\label{equ:MSEC}
\end{eqnarray}
Hence, both for kriging and conditional simulations, good approximations will be obtained by finding adequate covariance functions $C_1(\bh)$ such that 
$\Delta_{MSE}(\bx,C_0,C_1)$ is minimized. Regarding the MSE criterion for simulations, one usually has $MSE^s(\bx,C_1) \geq MSE^s(\bx,C_0,C_1)$, i.e. 
$\sigma^2_{k,C_1}(\bx) \geq MSE(\bx,C_1)$. So the HT approach normally improves over the T method in terms of $MSE^s$. On a total of more than a million simulations with varying covariance functions, covariance tapers, range parameters and kriging geometry, we only observed a handful of cases where $MSE(\bx,C_1)>\sigma^2_{k,C_1}(\bx) $. In all these cases, the differences between both values were extremely small. 

We observe that in Eq. \ref{equ:post} the $Z^*(\bx)$ term is the same for T and HT but the term $Z_s(\bx)-Z_s^*(\bx)$ is simulated with the true covariance $C_0$ for HT and with the plug-in covariance $C_1$ for T. Hence, HT should present local variability more similar to F than the T approach. \cite{stein2013} noted that the conditional variance of $Z_{cs}(\bx)$ obtained by HT  corresponds to the prediction error variance $MSE(\bx,C_1)$ obtained with the tapered covariance.

Since the taper function is a covariance function normalized such that $C_{T}(\bzero)=1$, the tapered covariance verifies $C_1(\bh)=C_0(\bh)C_T(\bh) \leq C_0(\bh)$. We therefore anticipate that the kriging variance can only increase when less spatial correlation is present. We establish the following result (see Appendix for a proof):
\begin{Proposition} 
\label{prop:prop3}
In the setting above, one has $\sigma^2_{k,C_1}(\bx) \geq \sigma^2_{k,C_0}(\bx)$
 for all $\bx \in D$.
\end{Proposition}
 
\subsection{Spectral densities of tapers and covariances}

From Bochner's theorem \citep{bochner1933monotone,chiles2012}, a stationary covariance function in $\Re^d$ has the spectral representation
\begin{equation}
C(\bh)=\int_{\Re^d} e^{i (\bomega\cdot \bh)} F(d\bomega),\quad \bh \in \Re^d
\label{equ:spectral}
\end{equation}
where  $F(d\bomega)$ is a positive bounded symmetric spectral measure on $\Re^d$ and $(\bh \cdot \bomega)$ denotes the inner product between vectors $\bh$ and $\bomega \in \Re^d$. When $C(\bh)$ is square integrable, the spectral measure can be written as a spectral density, $F(d\bomega)=f(\bomega)d\bomega$. Moreover, when $C(\bh)$ is absolutely integrable, the spectral density can be obtained by inversion of Eq. \ref{equ:spectral} \citep{lantuejoul_geostatistical_2002}:
\begin{equation}
f(\bomega)=\frac{1}{(2\pi)^d}\int_{\Re^d} e^{-i (\bomega\cdot \bh)} C(\bh) d\bh,
\quad \bomega \in \Re^d.
\label{equ:fourier_density}
\end{equation}
When $C(\bh)$ is isotropic, one can write $C(\bh) = \phi(\|\bh\|)$, where 
$\|\cdot\|$ denotes the Euclidean norm and $\phi$ is a continuous mapping $\phi: [0,\infty) \to \Re$ with $\phi(0) >0$. The previous equation simplifies to a 1D integral transform \citep{sneddon1951,lantuejoul_geostatistical_2002}:
\begin{equation}
f(\bomega)=\frac{1}{(2\pi)^{d/2}\|\bomega\|^{d/2-1}}\int_{0}^{\infty} J_{d/2-1}(\|\bomega\| r) \phi(r) r^{d/2} dr, \quad \bomega \in \Re^d
\label{equ:density_iso}
\end{equation}
where $J_\nu(\cdot)$ is the Bessel function of the first kind of order $\nu$.  

Tables \ref{tab:taper} and \ref{tab:covar} present some taper and covariance functions along with their spectral densities in $\Re^3$ obtained using Eq.~\ref{equ:density_iso}. For simplification, all functions are expressed as isotropic correlation functions with normalized distance $r=\|\bh\|/a$, and we denote $s=\|\bomega\|$ the norm of the frequency vector. With a slight abuse of notations, we shall write in the isotropic case $f(s)$ for $f(\bomega)$. Results can be extended to the anisotropic case following \cite{marcotte2015,marcotte2016tbm}.  
For the Cauchy covariance Eq.~\ref{equ:density_iso} could be solved only for $\alpha>1/2$ as in \cite{marcotte2015}. However, following an observation by \cite{lim2008}, the spectral density expression remains valid for $0<\alpha\le 1/2$ as can be checked by evaluating directly Eq. \ref{equ:spectral}. The smoothness of the covariance functions is $p-1$, where $p$ is the order of the lowest odd monomial in $r$. 
The decay rate of the spectral densities $f(s)$ at large frequencies is either exponential (Gaussian and Cauchy) or is  determined by the largest negative exponent on $s$. For the Cauchy spectral density, we use the fact that decay rate of Bessel functions ($K_{\nu}$) is  proportional to $s^{-1/2}e^{-s}$ for all $\nu$ \citep{lim2008}.

\begin{sidewaystable}[ht]
\small
\centering
\caption{Radial taper models ($r_\theta=r/\theta$; $I(r_\theta)=1$ if $r_\theta \le 1$, 0 otherwise), associated radial spectral density (with $s>0$) in $\Re^3$, smoothness of the taper and decay rate at large frequency $s$. The smoothness is the mean-square differentiability order of the associated random fields. 0 corresponds to mean-square continuity.}

\begin{tabular} {l l l c  c  }
\hline
Name & $\phi_\theta(r)$ & $f(s)$ &Smoothness & Decay rate\\
\hline \\[-3pt]
Spherical & $(1- \frac{3}{2} r_\theta+\frac{1}{2}r_\theta^3)\;I(r_\theta)$ &  $ \frac{3}{4\pi s^3} J_{3/2}^2(s/2)$ & 0 & $ s^{-4}$\\[4pt]
Cubic & $(1 - 7r_\theta^2 + \frac{35}{4}r_\theta^3 - \frac{7}{2}r_\theta^5 + \frac{3}{4} r_\theta^7)\;I(r_\theta)$   & $\frac{210 \left(\left(s^2-12\right) \sin \left(\frac{s}{2}\right)+6 s \cos \left(\frac{s}{2}\right)\right)^2}{\pi ^2 s^{10}} $ & 2 & $ s^{-6}$\\[4pt]
Penta & $(1 - \frac{22}{3}r_\theta^2 + 33r_\theta^4 - \frac{77}{2}r_\theta^5 + \frac{33}{2} r_\theta^7 - \frac{11}{2} r_\theta^9 + \frac{5}{6}r_\theta^{11})\;I(r_\theta)$ & $ \frac{27720 \left(s \left(s^2-60\right) \cos \left(\frac{s}{2}\right)-12 \left(s^2-10\right) \sin \left(\frac{s}{2}\right)\right)^2}{\pi ^2 s^{14}}$ & 4 & $ s^{-8}$\\[4pt]
Bonham & $ (1-r_\theta) \frac{\sin(2\pi r_\theta)}{2\pi r_\theta}+\frac{1-\cos(2\pi r_\theta)}{2\pi^2 r_\theta} $ & $\frac{4(1 - \cos(s)}{(s^3-4 \pi^2 s)^2} $ & 2 & $s^{-6}$ \\[4pt]
Wendland$_0$& $(1 - 2 r_\theta + r_\theta^2) \;I(r_\theta)$ &  $\frac{2 s-3 \sin (s)+ s \cos (s)}{\pi ^2 s^5}$ & 0 & $s^{-4}$\\[4pt]
Wendland$_1$& $(1 - 10 r_\theta^2 + 20 r_\theta^3 - 15 r_\theta^4 + 4 r_\theta^5) \;I(r_\theta)$ & $-\frac{60 \left(-4 s^2+\left(s^2-24\right) \cos (s)-9 s \sin (s)+24\right)}{\pi ^2 s^8}$ & 2 & $s^{-6}$\\[4pt]
Wendland$_2$& $(1 -\frac{28}{3} r_\theta^2 + 70 r_\theta^4 - \frac{448}{3} r_\theta^5 + 140 r_\theta^6 - 64 r_\theta^7 + \frac{35}{3} r_\theta^8) I(r_\theta)$ & $\frac{6720 \left(8 s \left(s^2-24\right)+9 \left(35-2 s^2\right) \sin (s)+s \left(s^2-123\right) \cos (s)\right)}{\pi ^2 s^{11}}$ & 4 & $s^{-8}$\\[4pt]
\hline
\end{tabular}
\label{tab:taper}
\end{sidewaystable}

\begin{sidewaystable}[ht]
\small
\centering
\caption{Isotropic covariance models $\phi(r)$, ($r$: normalized distance), spectral density $f(s)$ in $\Re^3$, smoothness of the covariance and spectral density decay rate at large frequency $s$.}
\begin{tabular} {l l l c c  }
\hline
Name & $\phi(r)$ &  $f(s)$ & Smoothness & Decay rate\\
\hline \\[-3pt]
Cauchy$_{\alpha}$ & $( 1 + r^2)^{-\alpha}$ with $\alpha>0$ & $\frac{2^{-\alpha -\frac{1}{2}} s^{\alpha -\frac{3}{2}} K_{\frac{3}{2}-\alpha }(s)}{\pi ^{3/2} \Gamma (\alpha )} $ & $\infty$ & $s^{\alpha-2}e^{-s}$ \\[4pt]
Matern$_{\nu}$ & $\frac{1}{2^{\nu-1}\Gamma(\nu)} r^\nu K_\nu(r)$ with $\nu > 0$ & $\frac{\left(s^2+1\right)^{-\nu -\frac{3}{2}} \Gamma \left(\nu +\frac{3}{2}\right)}{\pi ^{3/2} \Gamma (\nu )} $ & $2 \left\lceil \nu-1 \right\rceil$ & $s^{-(2\nu+3)}$\\[4pt]
Exponential & $e^{-r}$ & $\frac{1}{\pi ^2 (1+s^2)^2} $ & 0 & $s^{-4}$ \\[4pt]
Gaussian & $ e^{-r^2}$ & $\frac{e^{-\frac{s^2}{4}}}{8 \pi ^{3/2}} $ & $\infty$ & $e^{-s^2}$ \\[4pt]
\hline
\end{tabular}
\label{tab:covar}
\end{sidewaystable}

\subsection{Tail condition and convergence of HT for conditional simulations}
\label{sec:tail}

Let us consider $Z$, a zero mean second-order stationary Gaussian random field on $\Re^d$ with covariance function $C_0$, and let $C_1$ be another covariance function on $\Re^d$. Let us further denote their spectral densities $f_0$ and $f_1$, respectively. The tail condition is \citep{stein1993,stein1999,furrer2006} 
\begin{equation}
\stackrel{\lim} { \scriptstyle{s\rightarrow \infty}} \frac{f_1(s)}{f_0(s)}=\gamma,\quad 0 < \gamma < \infty.
\label{equ:tail3}
\end{equation}
When the tail condition is fulfilled, asymptotic equivalence of the mean squared prediction error for the two covariance functions can be established  under infill asymptotics (i.e. fixed domain, increasing number of observations, $n$) \citep{stein1993, stein1999,furrer2006}:
\begin{equation}
\stackrel{\lim} {\scriptstyle{n\rightarrow \infty}} \frac{MSE(\bx,C_1)}{MSE(\bx,C_0)}=1,  \qquad
\stackrel{\lim} {\scriptstyle{n\rightarrow \infty}} \frac{\sigma^2_{k,C_1}}{MSE(\bx,C_0)}=\gamma,
\label{equ:tail4}
\end{equation}
where it is recalled that in our context $C_1$ is the tapered covariance. Eq. (\ref{equ:tail4}) indicates that under infill asymptotics, the interpolation with $C_0$ or $C_1$ are asymptotically equivalent and that the ratio of the prediction variance tends to a non null, finite constant. In \cite{furrer2006} simulations show that the ratio $\sigma^2_{k,C_1}/MSE(\bx,C_0)$ stabilizes only slowly, especially for covariances with a more regular behavior at the origin. Note that the tail condition is a sufficient condition, but that it is not a necessary one. Establishing necessary conditions for Eq. (\ref{equ:tail4}) is still an open problem.
 Combining Eq. (\ref{equ:tail4}) with Eqs. (\ref{equ:MSEA}) to (\ref{equ:MSEC}) enables to state 
the following results.

\begin{Proposition} In the setting above,
\begin{equation}
\stackrel{\lim} {\scriptstyle{n\rightarrow \infty}} \frac{MSE^s(\bx,C_1)}{MSE^s(\bx,C_0)}=\frac{1}{2} \left(1+\frac{\sigma^2_{k,C_1}}{\sigma^2_{k,C_0}}\right) = \frac{1}{2}(1+\gamma)
\label{equ:tail5}
\end{equation}
and
\begin{equation}
\stackrel{\lim} {\scriptstyle{n\rightarrow \infty}} \frac{MSE^s(\bx,C_0,C_1)}{MSE^s(\bx,C_0)}=1.
\label{equ:tail6}
\end{equation}
\label{prop:prop4}
\end{Proposition}
Proposition \ref{prop:prop4} shows that the HT approach is asymptotically equivalent to F in terms of $MSE^s$ for simulation, but that the T approach is not. 

\medskip

The tail condition implies a similar decay rate of the spectral densities of taper, $C_T$, and the covariance function, $C_0$. The decay rate indicated in Tables \ref{tab:taper} and \ref{tab:covar} of covariance functions in $\Re^3$ enables to identify covariance-taper combinations that meet the tail condition, a sufficient condition to guarantee convergence of $MSE^s$ ratio with increasing $n$. Note the direct connection of decay rate and smoothness in these Tables. As an example the exponential covariance could be tapered with a spherical or a Wendland$_0$ taper as they all have the same decay rate. For the specific case of the Mat\'ern covariance, \cite{furrer2006} and \cite{zhang2008} proved that the taper function can present a faster decay rate than the Mat\'ern covariance. This result has been extended to other covariances by \cite{stein2013}. Note also that the tail condition is not fulfilled by any of the tapers considered for the Cauchy and the Gaussian covariances as they both show an exponential rate of decay of their associated spectral density. To the best of our knowledge, finding covariance functions defined on compact support with exponential rate of decay of their spectral density is still an open problem. 

\begin{figure}[ht]
\centering
\includegraphics[width=5.75cm]{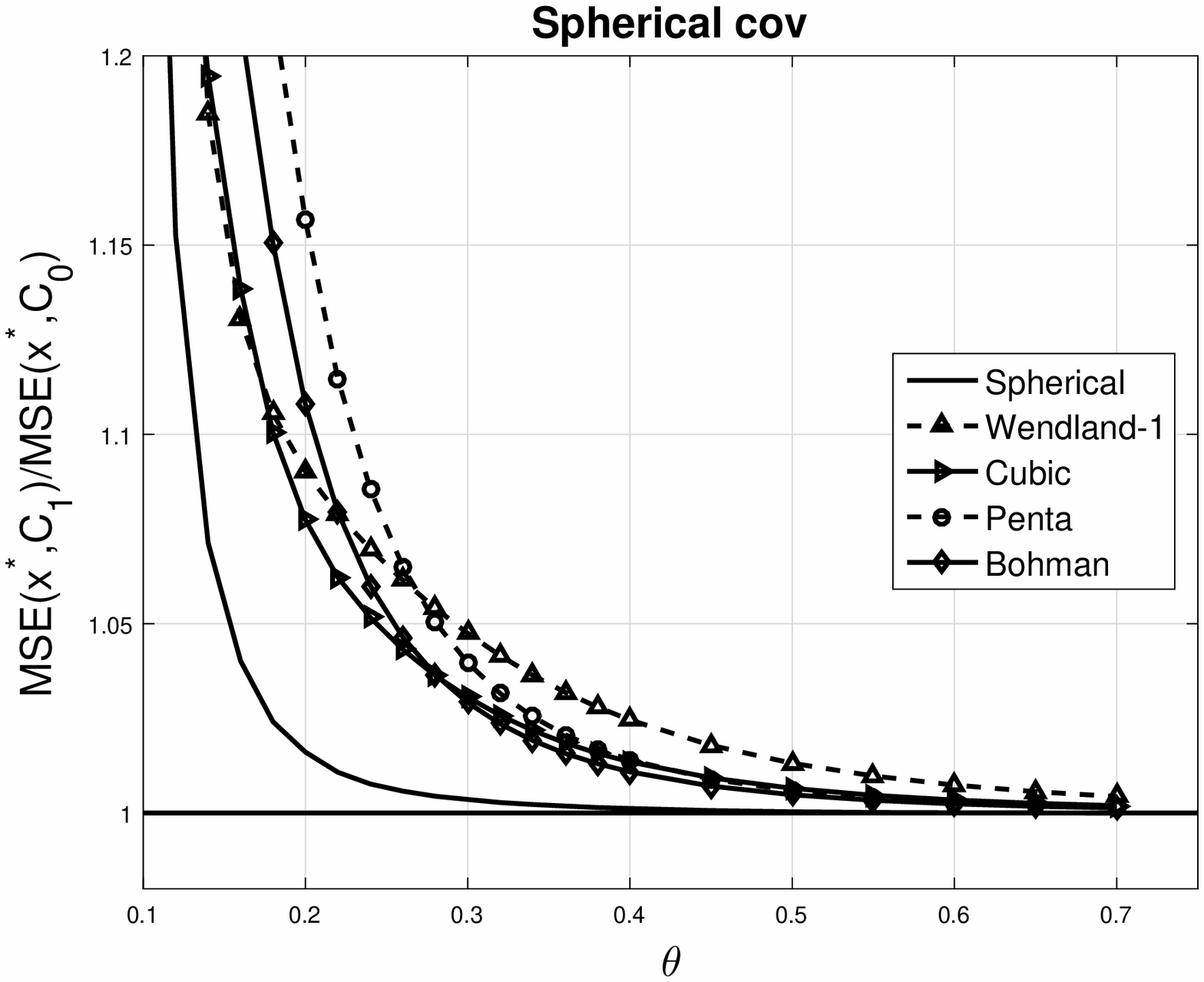} 
\includegraphics[width=5.75cm]{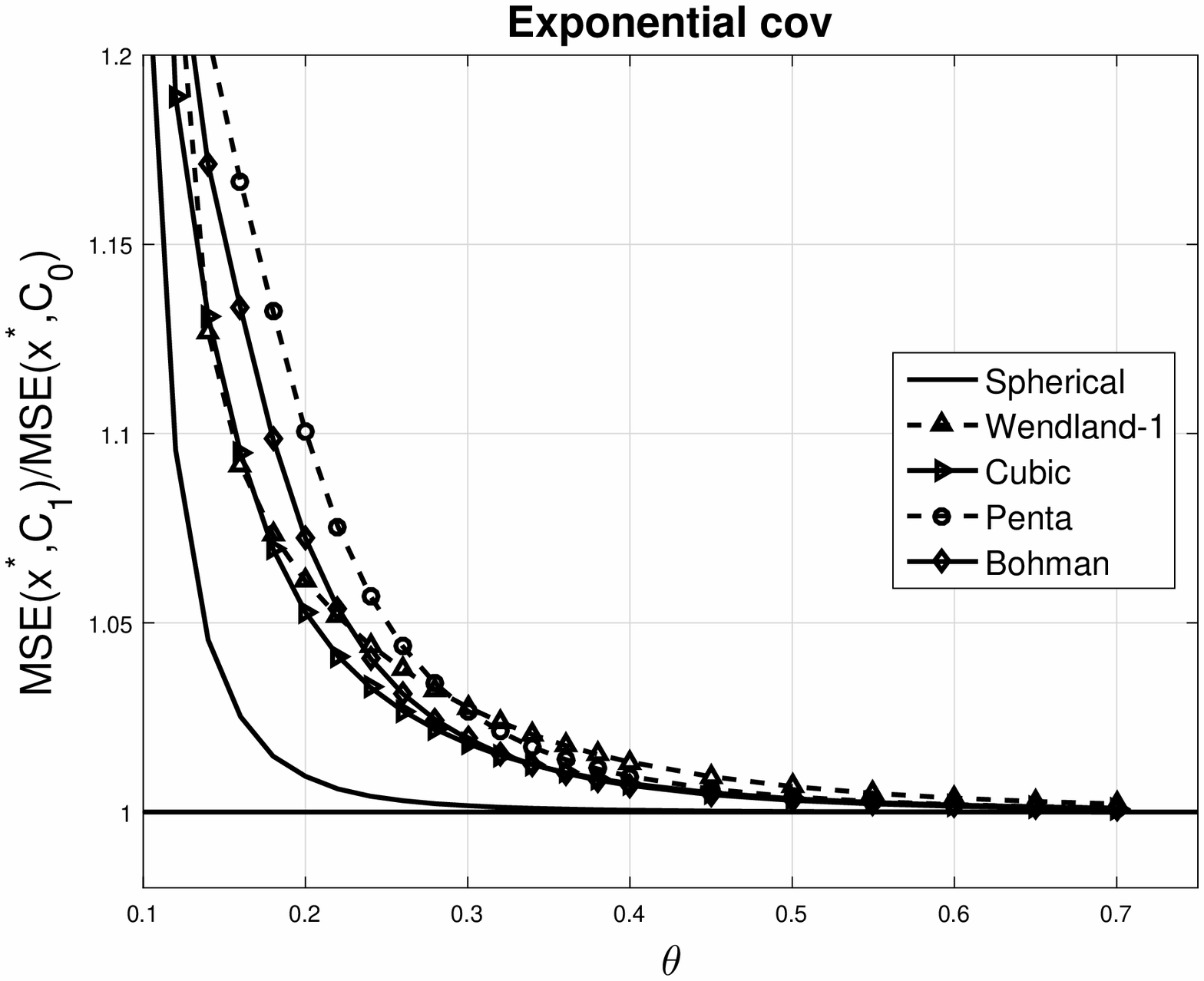} 
\includegraphics[width=5.75cm]{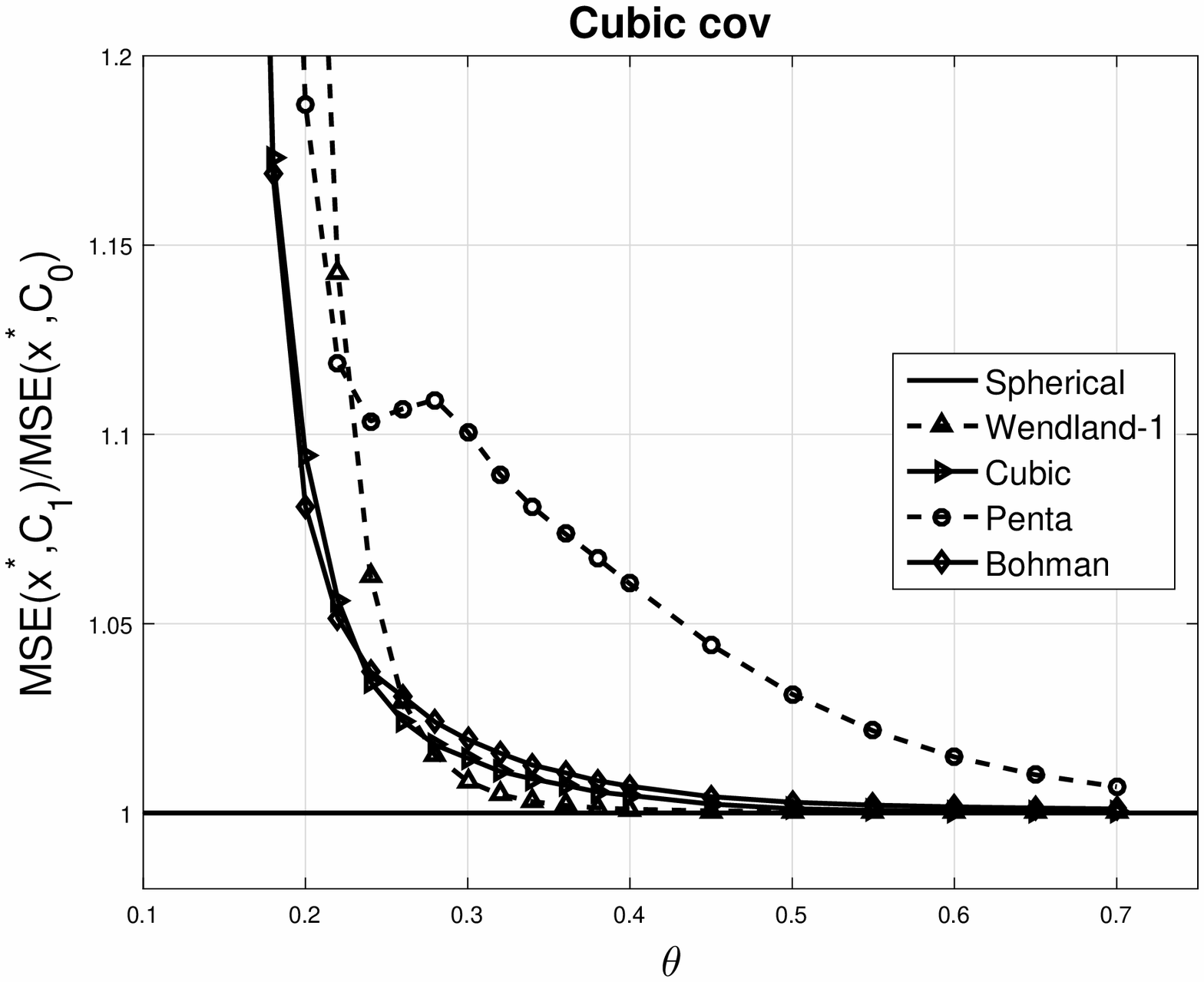}
\includegraphics[width=5.75cm]{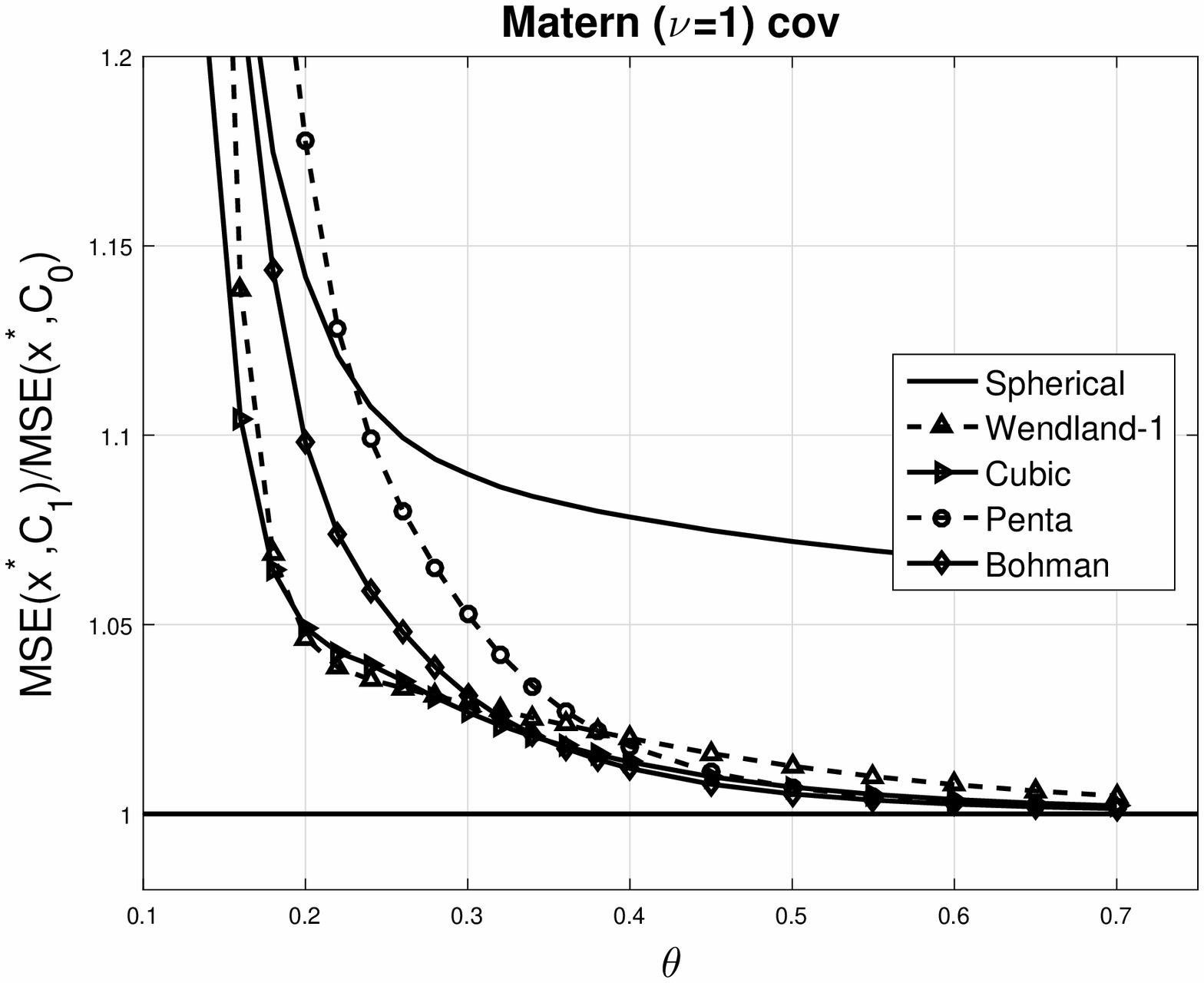}
\caption{Prediction MSE Ratio as a function of the range of the taper functions: spherical, cubic, penta, Bohman and Wendland$_1$. Covariance models in reading order: spherical, exponential, cubic, Matern with $\nu=1$. All covariances have unit range or unit effective range. MSE ratios computed over fifty random stratified samples of size $n=400$ for prediction at the center of the unit square (2D case).}
\label{fig:mse}
\end{figure}

Figure \ref{fig:mse} illustrates the prediction MSE ratios obtained as a function of the taper range, $\theta$, for a fixed sampling density. A random stratified sample of size 400 is taken over the unit square and the MSE is computed at the central point. This is repeated 50 times and the average MSE is retained to compute the MSE ratio. Performances of the tapers are measured by how short the range $\theta$ is for a given MSE ratio. Equivalently it can be measured by how close to 1 is the MSE ratio for a given taper range. When $C_0$ are spherical, exponential and cubic covariances, the best taper functions share the same mean squared differentiability and decay rate as $C_0$. When $C_0$ is a Mat\'ern covariance with $\nu=1$, the smoothness is ${\cal C}^0$, which corresponds to mean-squared continuity but not  mean-squared differentiability. The corresponding decay rate is $s^{-5}$, which is not achieved by any of the taper in Table \ref{tab:taper}. As expected, the ${\cal C}^2$ cubic and Wendland$_1$ tapers show better MSEs than the ${\cal C}^0$ spherical taper.   

\section{Degree of sparsity}
To help choosing the range of the taper function, it is useful to approximate the sparsity of the covariance matrix for the post-conditioning. For this, we assume the data points are randomly located within the field. Then, the distance between a pair of (distinct) points is also random. This distance is the one used to compute off-diagonal terms in the covariance matrix. The distance for the diagonal of the covariance matrix is obviously zero.
   
The cumulative distribution of the distance between two points randomly and uniformly located within an hypersphere with unit radius in $\Re^d$ has been studied in \cite{deltheil1926}. It is given by:
\begin{equation}
F_d(r)=r^d {\rm IB}_{1-r^2/4} \left (\frac{d+1}{2},\frac{1}{2} \right)+{\rm IB}_{r^2/4}\left(\frac{d+1}{2},\frac{d+1}{2}\right)
\label{equ:sparse}
\end{equation}
where $0 \leq r \leq 2$ and ${\rm IB}_z(a,b)$ stands for the incomplete beta function. 

The sparsity index of the tapered covariance matrix for points randomly located over the disk or sphere with unit radius can be assessed by taking:
\begin{equation}
S(\theta)=1-F_d(\theta)-(1-F_d(\theta))/n
\label{equ:sparse2}
\end{equation} 
where $\theta$ is the finite range of the tapering function and $n$ is the number of samples within the disk/sphere of unit radius. The term $(1-F_d(\theta))/n$ takes into account the diagonal of the covariance matrix that reduces the overall sparsity. For $n$ large this term becomes negligible and one has $S(\theta)\approx 1-F_d(\theta)$. 

\begin{figure}[ht]
\centering
\includegraphics[width=5.75cm]{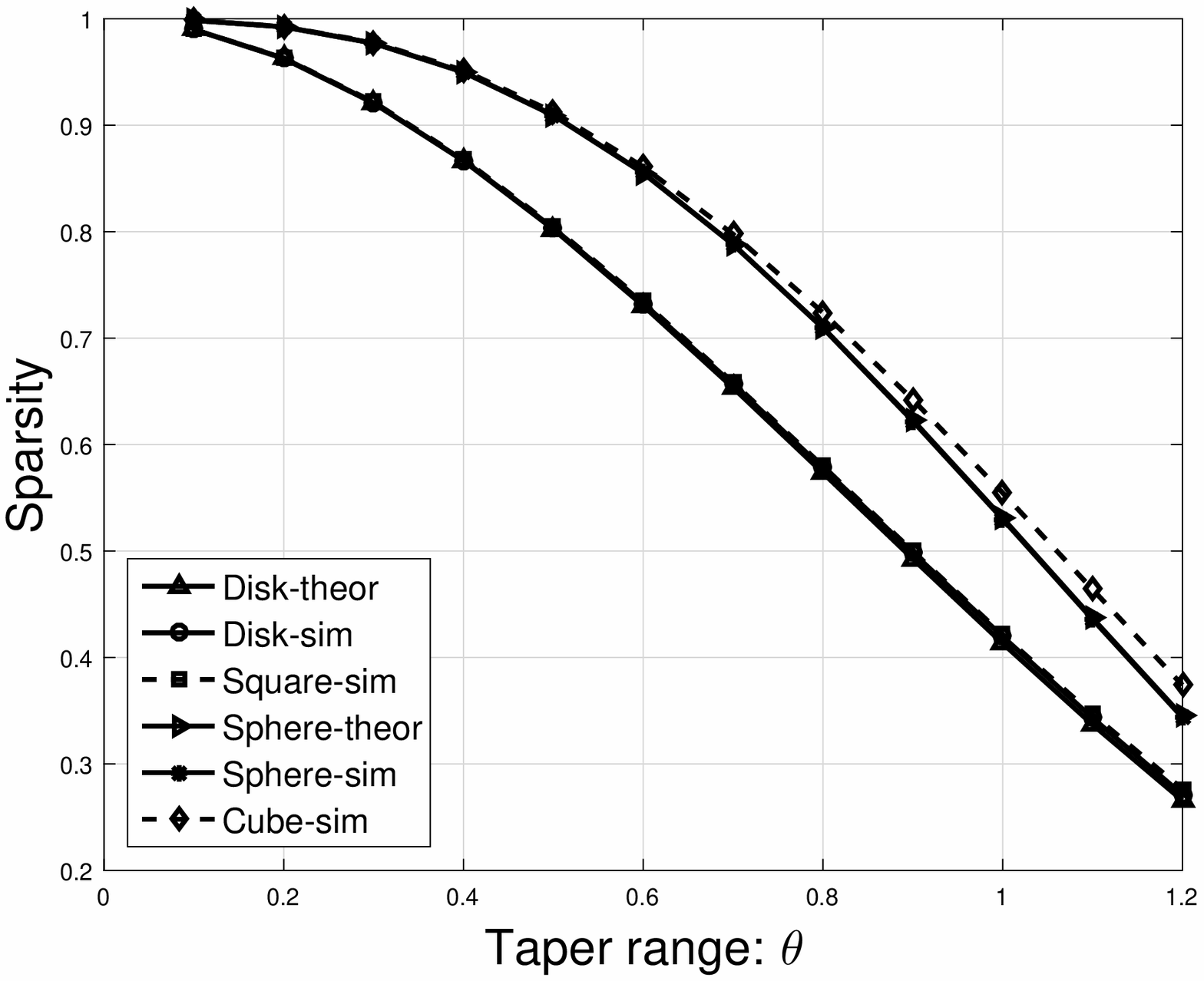} 
\includegraphics[width=5.75cm]{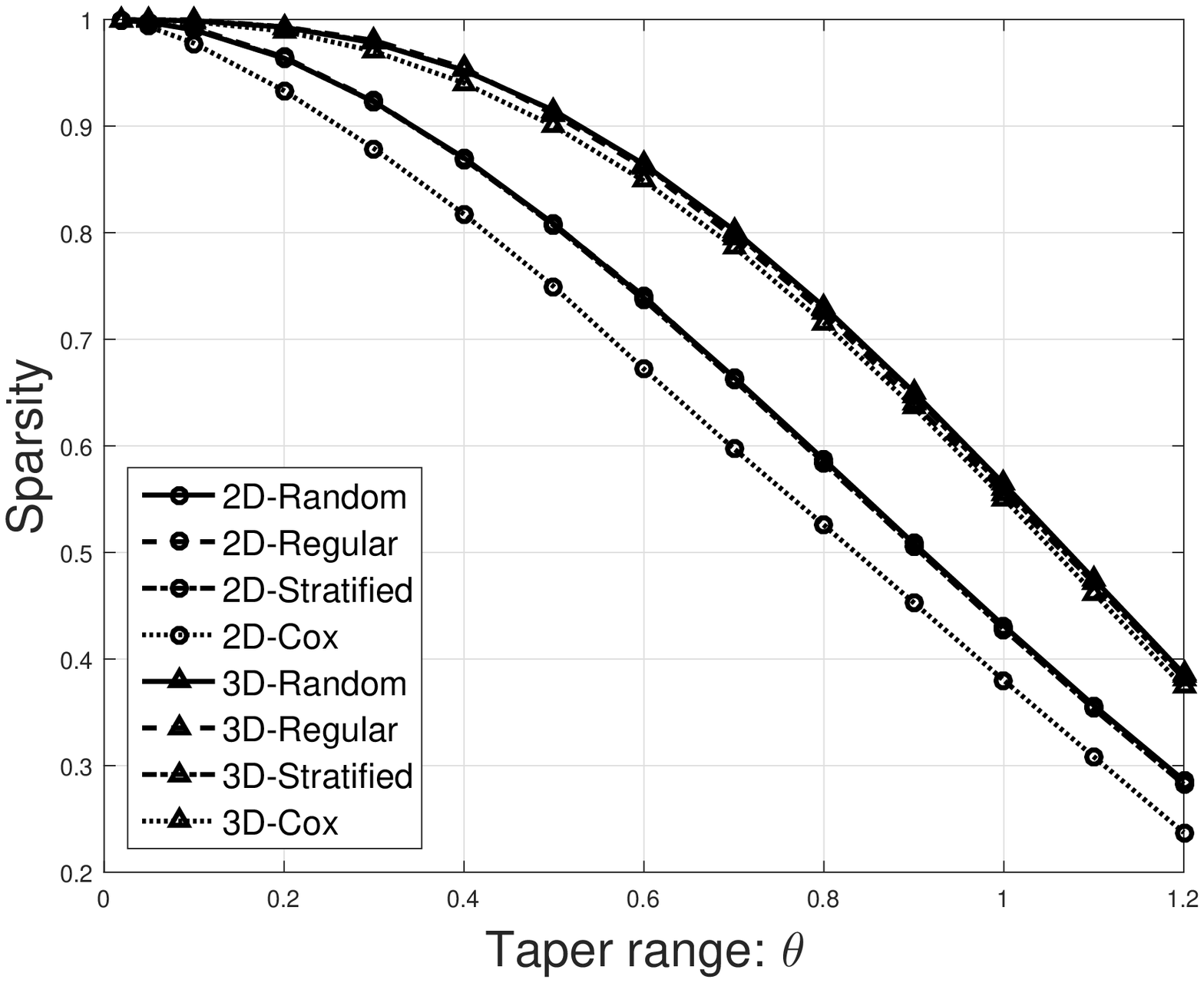}
\caption{Left: Theoretical and experimental sparsity of the covariance matrix for 2000 random points over unit radius disk or sphere, and a square or cube of the same size, as a function of the tapering range.  Right: Experimental sparsity over a square or cube for four different sampling designs, three uniform (random, regular and random stratified) with 50 x 50 points (2D) and 15 x 15 x 15 points (3D) and one clustered (Cox process with same number of points)}
\label{fig:sparsity}
\end{figure}

Cumulative distribution of distance for points randomly located over rectangles and cube can be found in the literature \citep{philip1991}. However, the relations are rather lengthy and cumbersome to use. Moreover, the distribution of data points is rarely truly random in practice. As we only need an approximation of the sparsity, we instead assume that the sparsity for a square or a cube can be well approximated by a disk or a sphere with the same area/volume and uniform random location within. Figure \ref{fig:sparsity} confirms the validity of this assumption by comparing the theoretical sparsity index obtained with Eq. \ref{equ:sparse2} and the sparsity obtained experimentally after covariance tapering with a finite range of $\theta$ for a simulation with 2000 points randomly located within a square of area $\pi$ and a cube of volume $4\pi/3$. The approximation with the disk/sphere for the square/cube is deemed sufficiently precise for our purpose. Discrepancies appear non-negligible only at small sparsity values where tapering is anyway not useful. The figure also shows that larger sparsity is expected for the 3D case than for the 2D case. However, even in the 2D case, sparsity  $>0.9$ can be obtained for $\theta\leq 0.3$. 

Hence, for a rectangular area of size $a \times b$ and a taper function with finite range $d$, the sparsity can be approximated by computing first $\theta=\frac{d\sqrt{\pi}}{\sqrt{a\cdot b}}$ and then evaluating Eq. \ref{equ:sparse2}. Similarly, in 3D for a cube of edges $a,b,c$ one computes $\theta=\frac{d(4\pi/3 )^{1/3}}{(a\cdot b \cdot c)^{1/3}}$ and then evaluates Eq. \ref{equ:sparse2}. As examples, suppose the taper range is selected as 0.6 times the effective range of the desired covariance and that the effective range represents $1/5$ of the side length of the square field under study. One computes $\theta=0.6 \times 0.2\sqrt{\pi}\approx 0.21$, $F_2(0.21) \approx 0.04$ and $S(\theta) \approx 96\%$. In 3D, one computes $\theta=0.6 \times 0.2 \times (4\pi/3)^{1/3}\approx 0.13$, $F_3(0.13) \approx 0.002$ and $S(\theta) \approx 99.8\%$. Hence, in 3D when $n=1 \times 10^5$, the covariance matrix required for conditioning by kriging using all data at once will have approximately $2\times 10^7$ non-zero entries. This is manageable as opposed to the prohibitive $10^{10}$ non-zeros entries for the full  covariance matrix. 

Figure \ref{fig:sparsity}-right compares the experimental sparsity of the covariance matrix obtained for samples taken following four sampling designs: i. regular grid,  ii. random stratified  (one random sample per cell), iii. purely random and iv. following a log-Gaussian Cox process as in \cite{bolin2016}. For the Cox process, locations are drawn from a Poisson process of intensity $\lambda( \bx ) = \exp \{ Z ( \bx )\}$ where $Z ( \bx )$ is a zero mean unit variance Gaussian field with exponential covariance and practical range 0.3 $\times$ the square or cube side length. Figure \ref{fig:sparsity} shows that the sparsity is robust to the exact sampling strategy as long as homogeneous coverage is ensured. However, less sparsity is obtained (especially in 2D) with the clustered sampling generated by the Cox process. In this case, it might be useful to adapt the taper range to the local sampling density as suggested by \cite{bolin2016}.

\section{Results}

The aim of this Section is to illustrate the ability of the HT approach to approximate accurately highly non linear characteristics of the simulated random fields, as compared to the much more computer intensive F approach. On four selected situations in 1D, 2D and 3D, we compare the conditional simulations obtained using three different approaches: i. the unconditional simulation and the conditioning by kriging are both done with the desired covariance (F), ii. the unconditional simulation and conditioning by kriging are both done with the tapered covariance (T), and iii. the unconditional simulation is done with the desired covariance and the conditioning by kriging is done with the tapered covariance (HT). Without loss of generality, small scale illustrative examples are used throughout to limit the computational burden of some response functions.

\subsection{1D example}
Two different forward models are computed on each of the realizations for the three approaches. The first forward model is the maximum absolute difference between consecutive points along the profile (discrete estimate of the slope) The second forward model is the total length of the 1D profile as in \cite{chiles2012}. The distributions of the responses are then compared using boxplots and two-sample Kolmogorov-Smirnov goodness of fit tests are computed. The p-values of the test statistic measure the similarity of the distributions. 

The profiles are simulated at 100 regularly spaced points.  The desired covariance, $C_0$, is exponential with effective range equal to 1/3 of the simulated length. Five hundred parent unconditional realizations are simulated. In each one a different sample is obtained by taking point 1, point 100 and eight points randomly picked among points 2 to 99. Points 1 and 100 are forced in the sample to avoid being in extrapolation mode. Each sample is then used to create a single conditional realization, so all conditional realizations are independent. All the simulations are done by Cholesky decomposition. As a consequence of the discussion in Section 2, the taper covariance function is set to be spherical. Its range $\theta$ is varied between 0.25 and 3 times the effective range. 

\begin{figure}[ht]
\centering
\includegraphics[width=11.5cm]{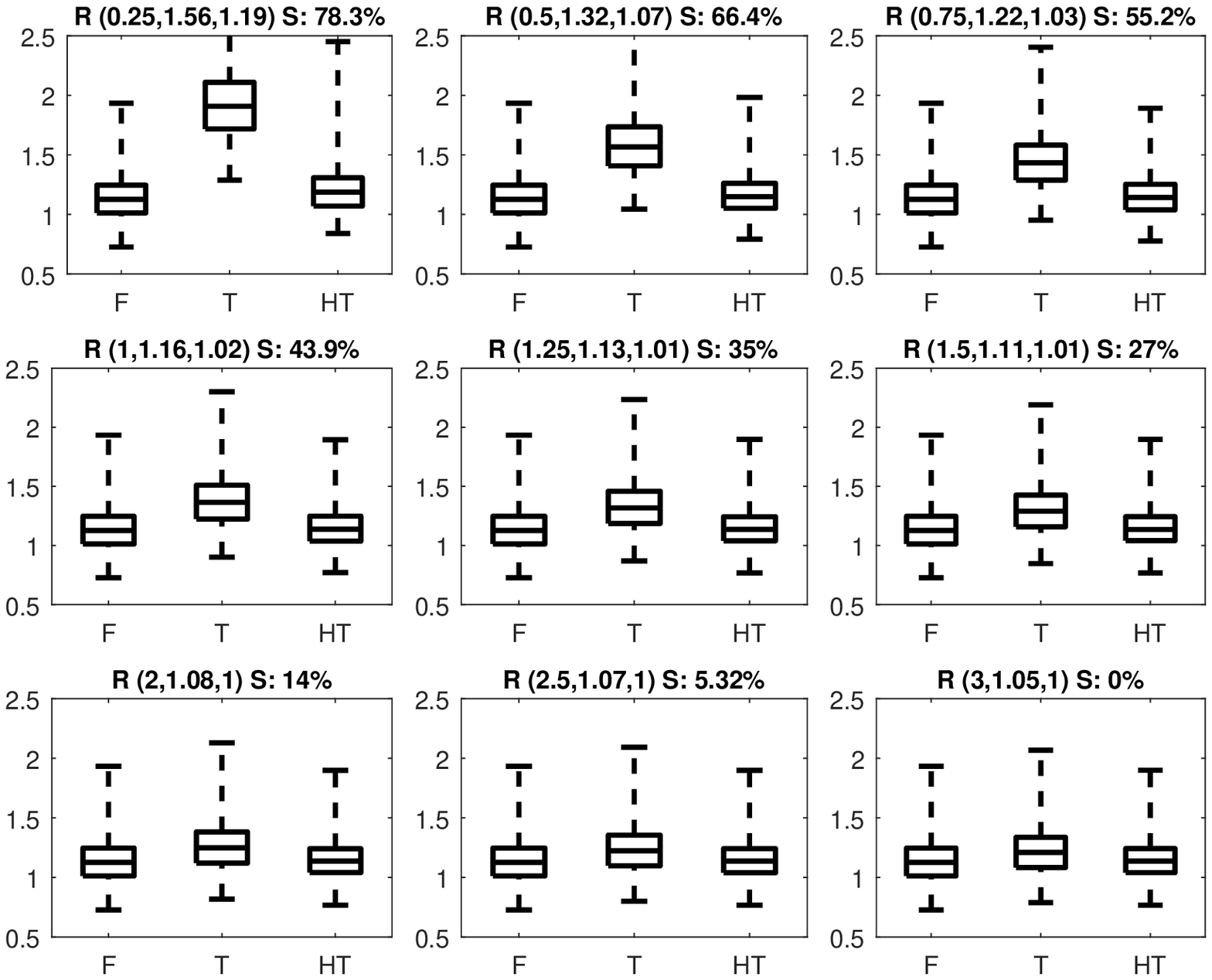} 
\caption{Boxplots of distributions of maximum absolute difference between consecutive points with F, T and HT. Ratios in the subtitles are: (taper range/effective range, $\frac{MSE^s(x,C_1)}{MSE^s(x,C_0)}$, $\frac{MSE^s(x,C_0,C_1)}{MSE^s(x,C_0)}$). Sparsity (S) is indicated. Desired covariance is exponential and taper covariance is spherical. Five-hundred independent conditional realizations with 10 conditioning points.}
\label{fig:taper_exp_spher_max}
\end{figure}

\begin{figure}[ht]
\centering
\includegraphics[width=8cm]{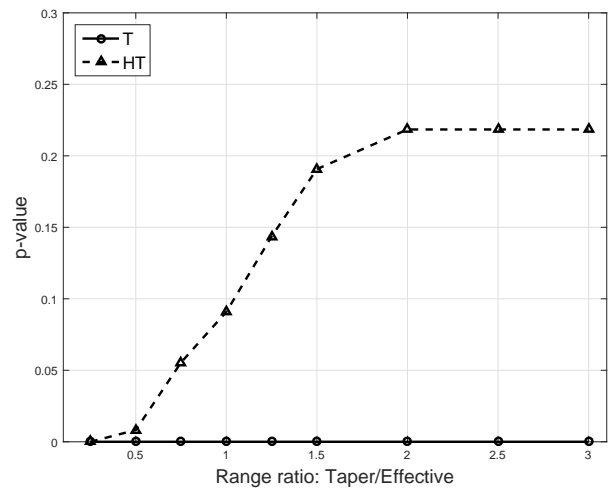} 
\caption{KS test p-values for equality of distributions of maximum absolute difference between consecutive points for T vs F and HT vs F. }
\label{fig:taper_exp_spher_max_pval}
\end{figure}

Figure \ref{fig:taper_exp_spher_max} presents  the boxplots for the three approaches when the response function is the maximum absolute difference between consecutive points. HT outperforms T as confirmed by the KS tests which indicate HT vs F distributions are not significantly different (at $\alpha=0.05$) for taper ranges above 0.75 times the effective range while they are significantly different in all cases for the T vs F approaches (see Fig. \ref{fig:taper_exp_spher_max_pval}).

\begin{figure}[ht]
\centering
\includegraphics[width=11.5cm]{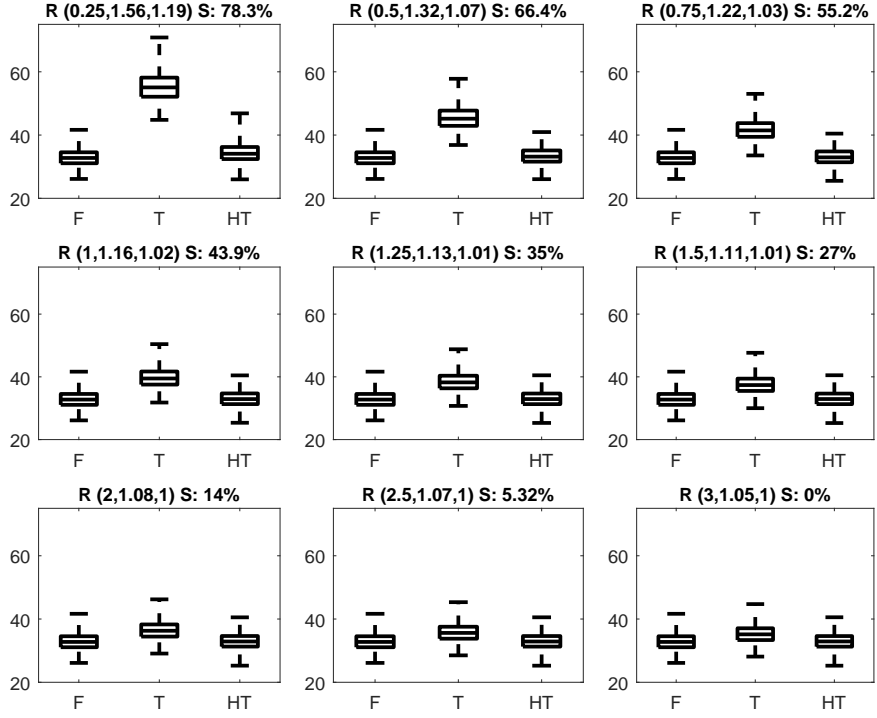} 
\caption{Boxplots of distributions of profile lengths (along ordinate) obtained with F, T and HT. Ratios in the subtitles are: (taper range/effective range, $\frac{MSE^s(x,C_1)}{MSE^s(x,C_0)}$, $\frac{MSE^s(x,C_0,C_1)}{MSE^s(x,C_0)}$). Sparsity (S) is indicated. Desired covariance is exponential and taper covariance is spherical. Five-hundred independent conditional realizations with 10 conditioning points.}
\label{fig:taper_exp_spher}
\end{figure}

\begin{figure}[ht]
\centering
\includegraphics[width=8cm]{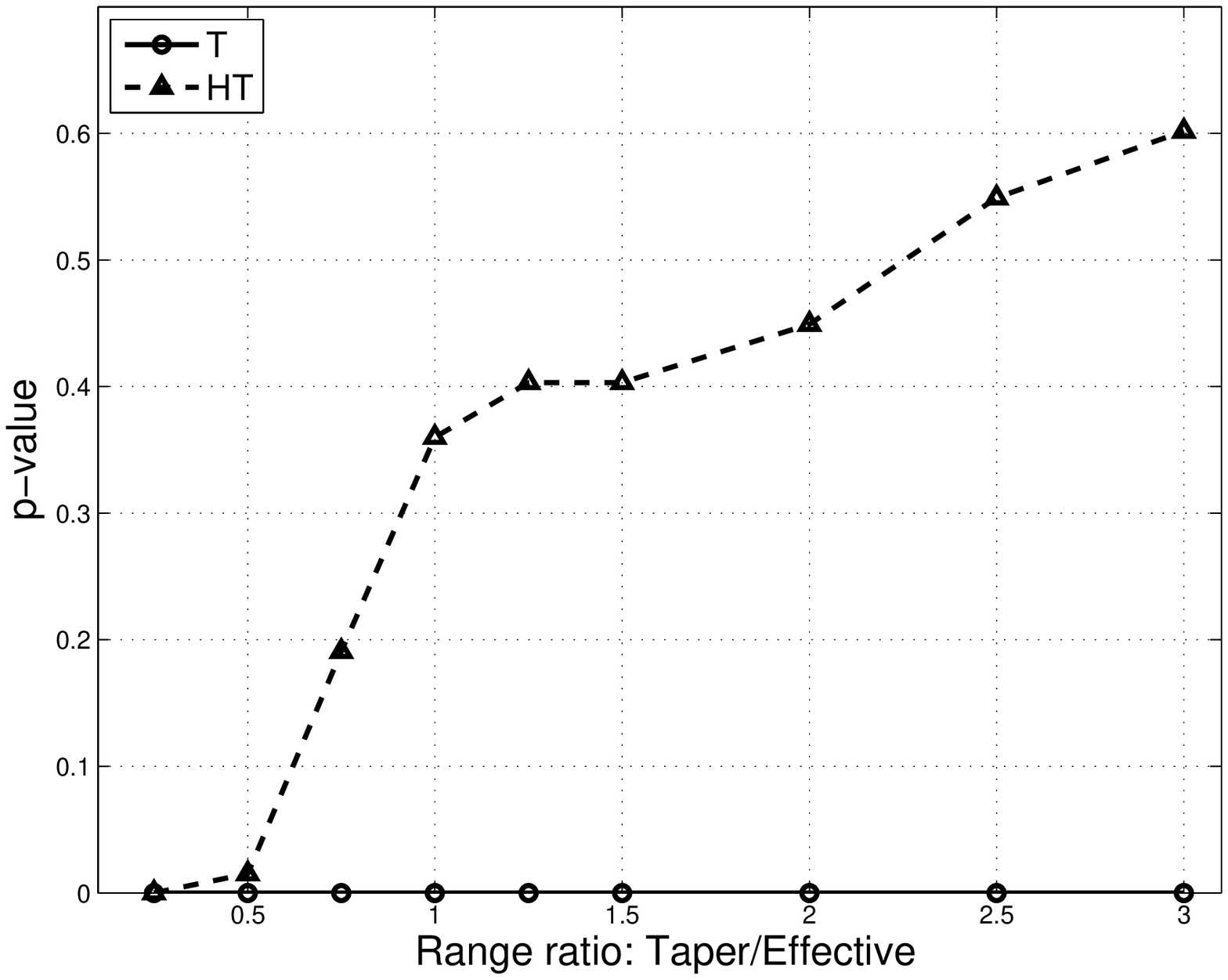} 
\caption{KS test p-values for equality of distributions of profile lengths for T vs F and HT vs F.}
\label{fig:taper_exp_spher_pval}
\end{figure}

Figure \ref{fig:taper_exp_spher} presents the boxplots for the three approaches for the response length of the profile. Clearly, the HT length distributions are more similar to the distributions obtained with F than the distributions with the T approach. This is again confirmed by the two-sample Kolmogorov-Smirnov (KS) test p-values (Fig. \ref{fig:taper_exp_spher_pval}) which are larger for the HT approach than for the T approach. In the latter case, significant differences (i.e. p-value$<0.05$) are obtained even for a taper range that is three times the effective range of the desired covariance. On the contrary, the HT approach indicates non-significant differences for a taper range representing 0.75 times the effective range. 

\subsection{2D examples}

In petroleum and environmental applications it is important that the simulated fields represent adequately the connectivity of facies. Moreover, flow path and corresponding transit time between two given sites are relevant for example to environmental contaminant transport problems or for assessment of potential localization of injection wells in reservoir engineering. It is thus of primary importance that tapering strategies reproduce accurately connectivity and flow characteristics.   In these experiments, $C_0$ is Mat\'ern with $\nu=1$ and  effective range equal to half the simulated length, and $C_T$ is Wendland$_1$, as shown in Section \ref{sec:tail} and illustrated in Figure \ref{fig:mse}. Forty samples of size 100 are taken as conditioning data from as many different unconditional realizations. For each sample, a series of 40 conditional realizations are produced for each approach F, T and HT. 

\subsubsection{Facies connectivity}
\begin{figure}[ht]
\centering
\includegraphics[width=10cm]{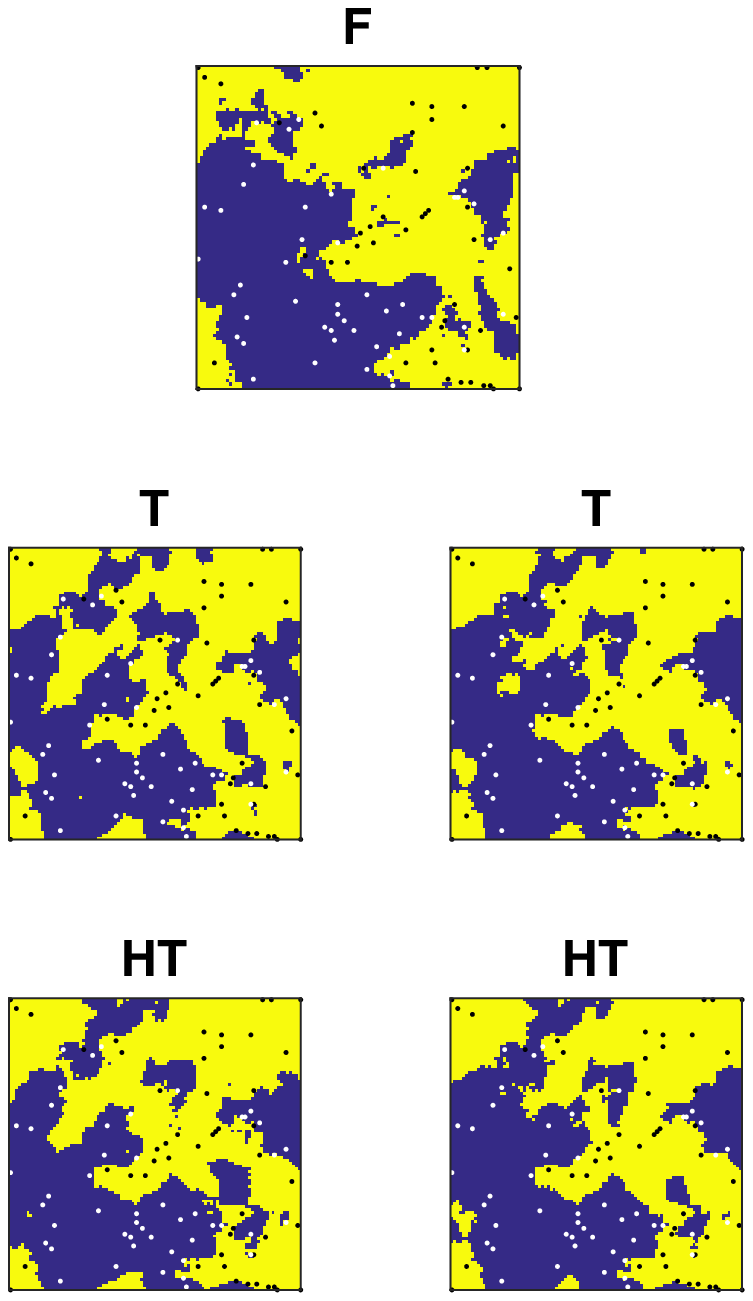} 
\caption{Realizations for approaches F (top row), T (middle row) and HT (bottom row) for ratios of taper range/effective range of 0.5 (left column) and 1.0 (right column). Desired covariance is Matern($\nu=1$), taper covariance is Wendland$_1$. Conditioning data location indicated by dots in each sub-figure.}
\label{fig:realizations_2d_allard}
\end{figure}

We first consider a two facies case: one permeable  and one impervious. \cite{renard2013} provide a review of the different metrics, both global and local, that can be used to measure connectivity. Among the global metrics, a useful one is the proportion of connected pairs among all pairs within the permeable facies \citep{hovadik2007}. More precisely, one computes:
\begin{equation}
g(p)=\frac{1}{n_p^2} \sum_{i=1}^{M}n_i^2
\label{equ:connect}
\end{equation} 
where $p$ is the proportion of permeable facies, $M$ is the number of clusters of permeable facies, $n_i$ is the size of permeable cluster $i$ and $n_p=\sum_{i=1}^{M}n_i$ is the total number of permeable cells. For a given proportion $p$, the permeable phase is defined as the set of points $\bx$  such that $Z(\bx)  \leq \Phi^{-1}(p)$, where $\Phi(\cdot)$ is the cumulative distribution function of a Gaussian $(0,1)$ random variable.  Specifically, a threshold is applied on the Gaussian realizations so as to get $p=0.3$ for each ensemble of realizations of F, T and HT. The connectivity measure $g(p)$ is then computed on these 2D truncated Gaussian random fields.   For illustration, Fig. \ref{fig:realizations_2d_allard} presents one randomly chosen conditional realization of the field by the F, T and HT methods for two different taper ranges. The same conditioning data is used in each case. Note that the four corners of the 2D field are always included in the conditioning samples again to avoid too strong extrapolation. 

\begin{figure}[ht]
\centering
\includegraphics[width=11.5cm]{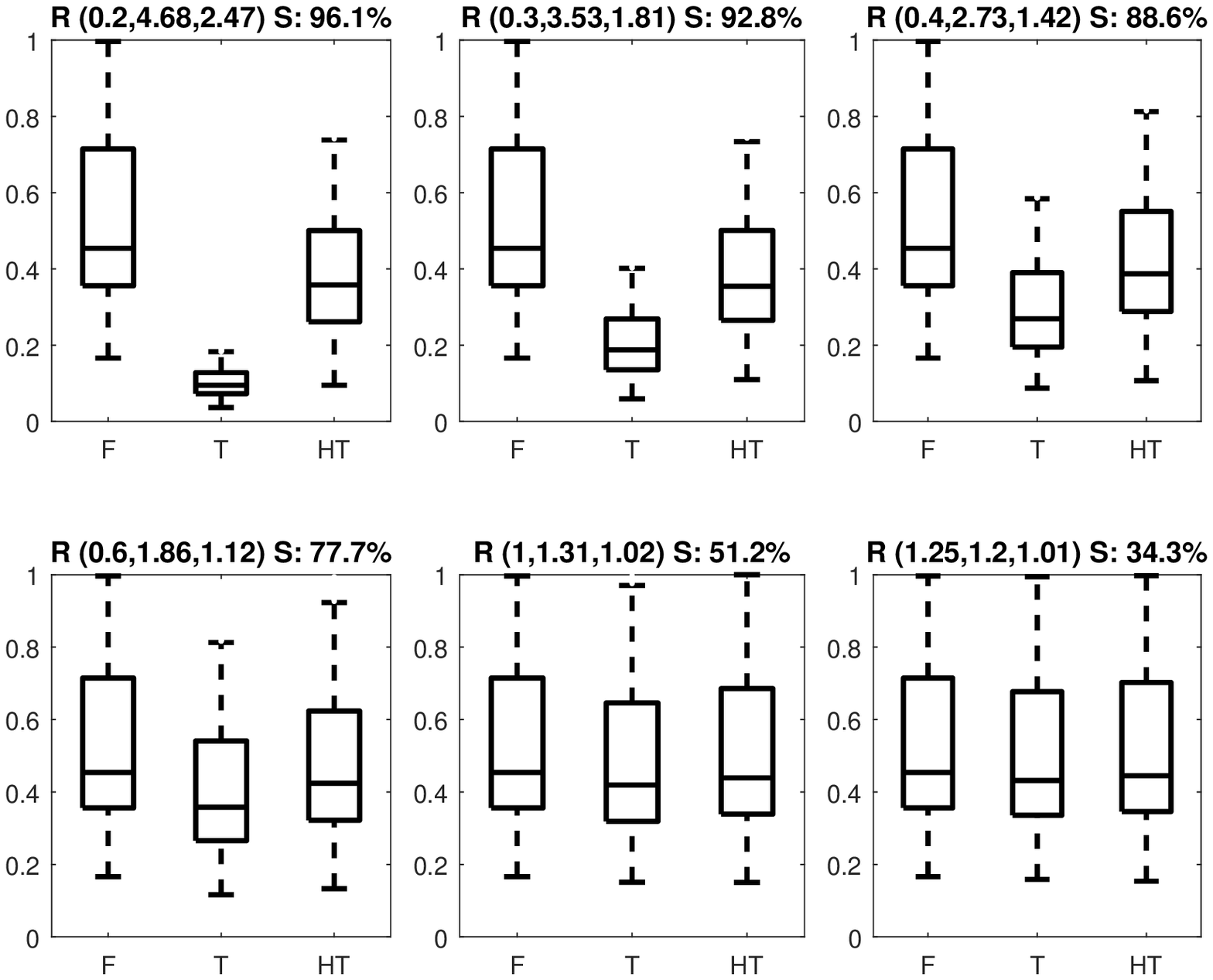} 
\caption{Boxplots of the 2D connectivity measure at $g(0.3)$ for approaches F, T and HT and for various ratios of taper range/effective range. Ratios in the subtitles are: (taper range/effective range, $\frac{MSE^s(x,C_1)}{MSE^s(x,C_0)}$, $\frac{MSE^s(x,C_0,C_1)}{MSE^s(x,C_0)}$). Desired covariance is Matern($\nu=1$), taper covariance is Wendland$_1$. Forty conditional realizations per sample of size 100 conditioning points. Results are accumulated over forty different samples for a total of 1600 values for each box.}
\label{fig:taper_mat_wen_2d_allard}
\end{figure}

Figure \ref{fig:taper_mat_wen_2d_allard} shows the boxplots of the connectivity measure for different values of the ratio taper range/effective range. Results show clearly that the T approach dramatically underestimates the connectivity measure, whereas it is reasonably well reproduced with the HT approach, even for relatively high  sparsity.

\subsubsection{Transit time}
\begin{figure}[ht]
\centering
\includegraphics[width=10cm]{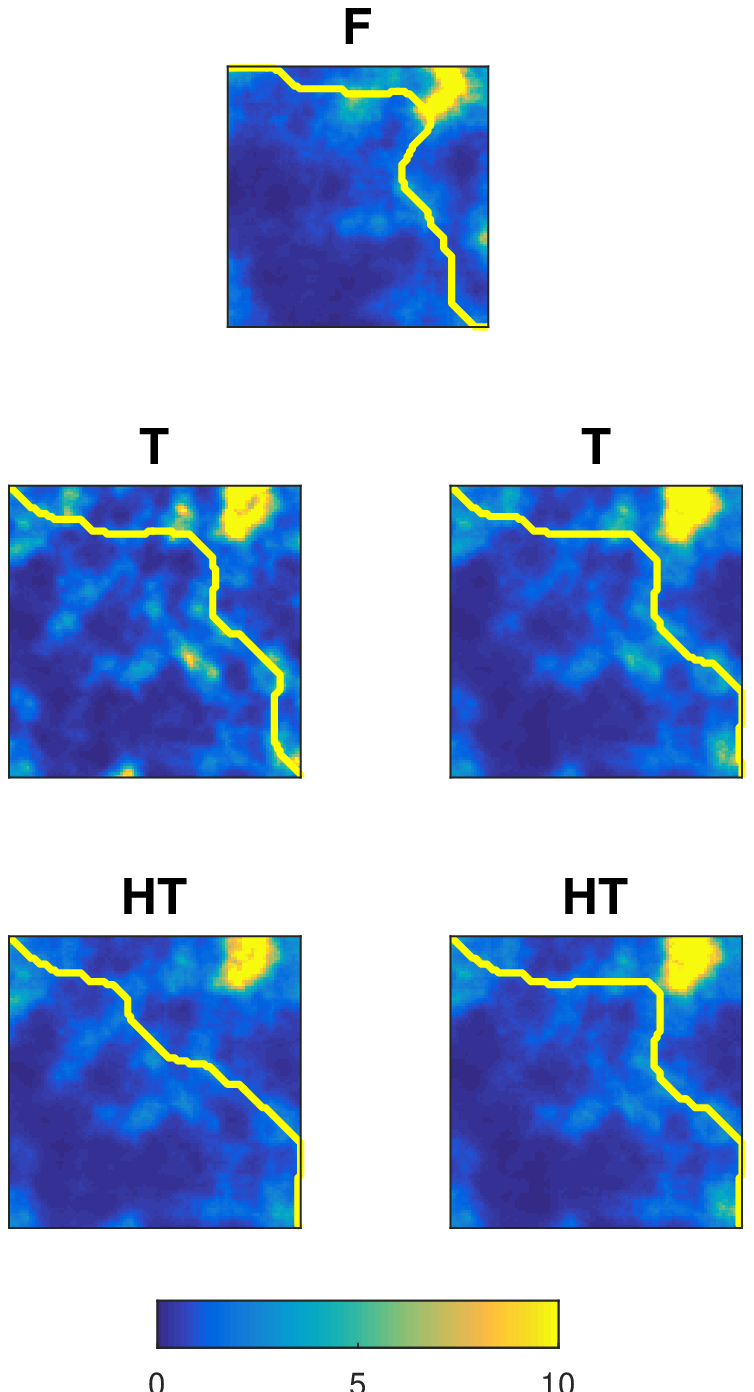} 
\caption{Realizations for approaches F (top row), T (middle row) and HT (bottom row) and for ratios of taper range/effective range of 0.5 (left column) and 1.0 (right column). Desired covariance is Matern($\nu=1$), taper covariance is Wendland$_1$. Transit times are respectively 184 for F, 148 and 165 for T, 165 and 175 for HT. Shortest path indicated by the yellow line.}
\label{fig:realizations_2dtemps_coin}
\end{figure}

\begin{figure}[ht]
\centering
\includegraphics[width=11.5cm]{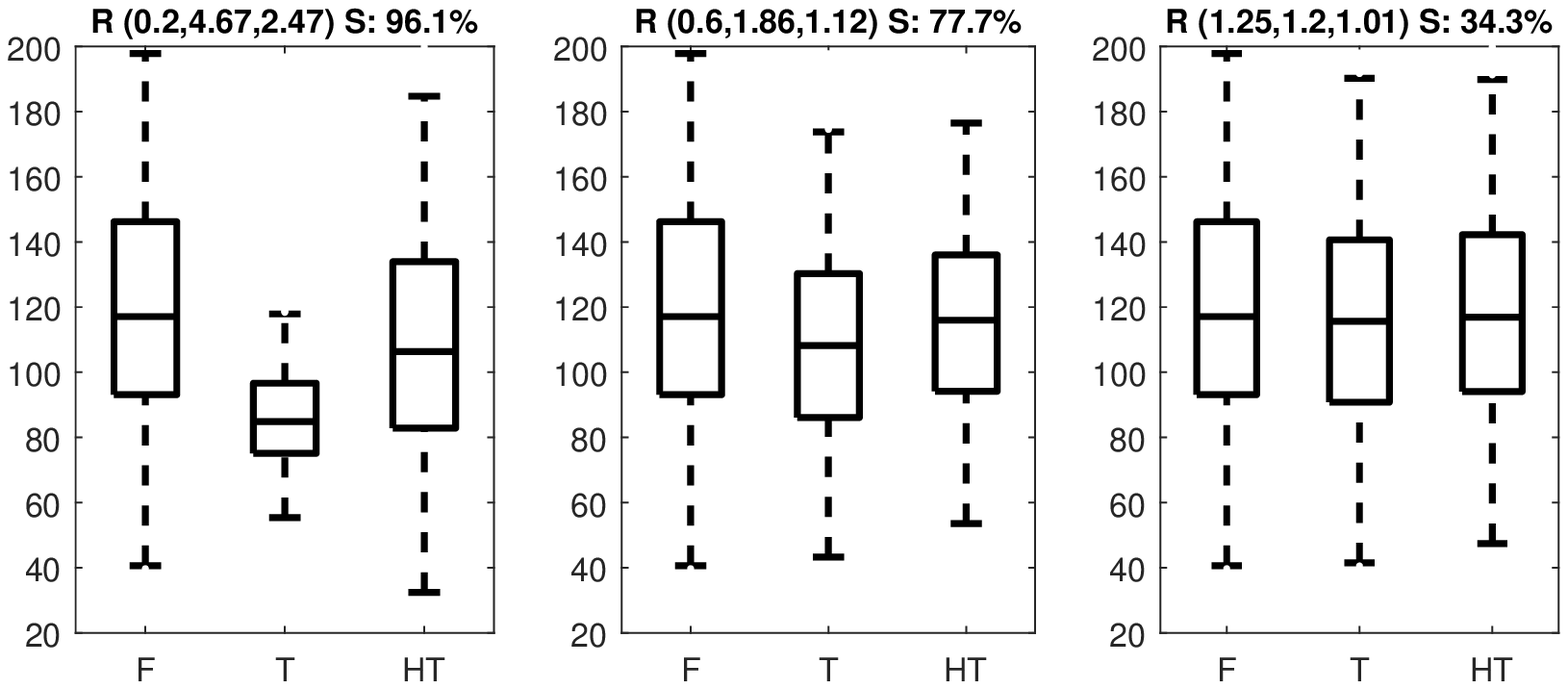} 
\caption{Boxplots of the transit times for approaches F, T and HT and for various ratios of taper range/effective range. Ratios in the subtitles are: (taper range/effective range, $\frac{MSE^s(x,C_1)}{MSE^s(x,C_0)}$, $\frac{MSE^s(x,C_0,C_1)}{MSE^s(x,C_0)})$. Desired covariance is Matern($\nu=1$) and taper covariance is Wendland$_1$. Ten conditional realizations per sample of size 100 conditioning points. Results are accumulated over 40 independent samples for a total of 400 values for each box.}
\label{fig:taper_mat_wen_2dtemps_coin}
\end{figure}

For the same set of simulations, a second response function is obtained by computing fastest transit time between the upper left cell and the bottom right cell of each realization. The exponential of the Gaussian fields $Z(\bx)$ are interpreted as speed. They are converted to slowness by taking the inverse of the speed. Then, the fastest path between the upper left point and the bottom right point of the field is determined by applying Dyjkstra algorithm \citep{dyjkstra1959}. Since transit times are rather slow to compute, we limit ourselves to 10 different conditional realizations for each of the 40 samples. Fig. \ref{fig:realizations_2dtemps_coin} shows examples of realizations obtained by the methods F, T and HT for two different taper ranges. The fastest path is also indicated.  The distributions of the $40 \times 10$ responses are described by boxplots for different taper ranges (see Fig. \ref{fig:taper_mat_wen_2dtemps_coin}). Again, the HT approach shows an excellent reproduction of the transit time, whereas the T approach requires a sparsity less 
than 77\%.
 
\subsection{A 3D example}

\begin{figure}[ht]
\centering
\includegraphics[width=11.5cm]{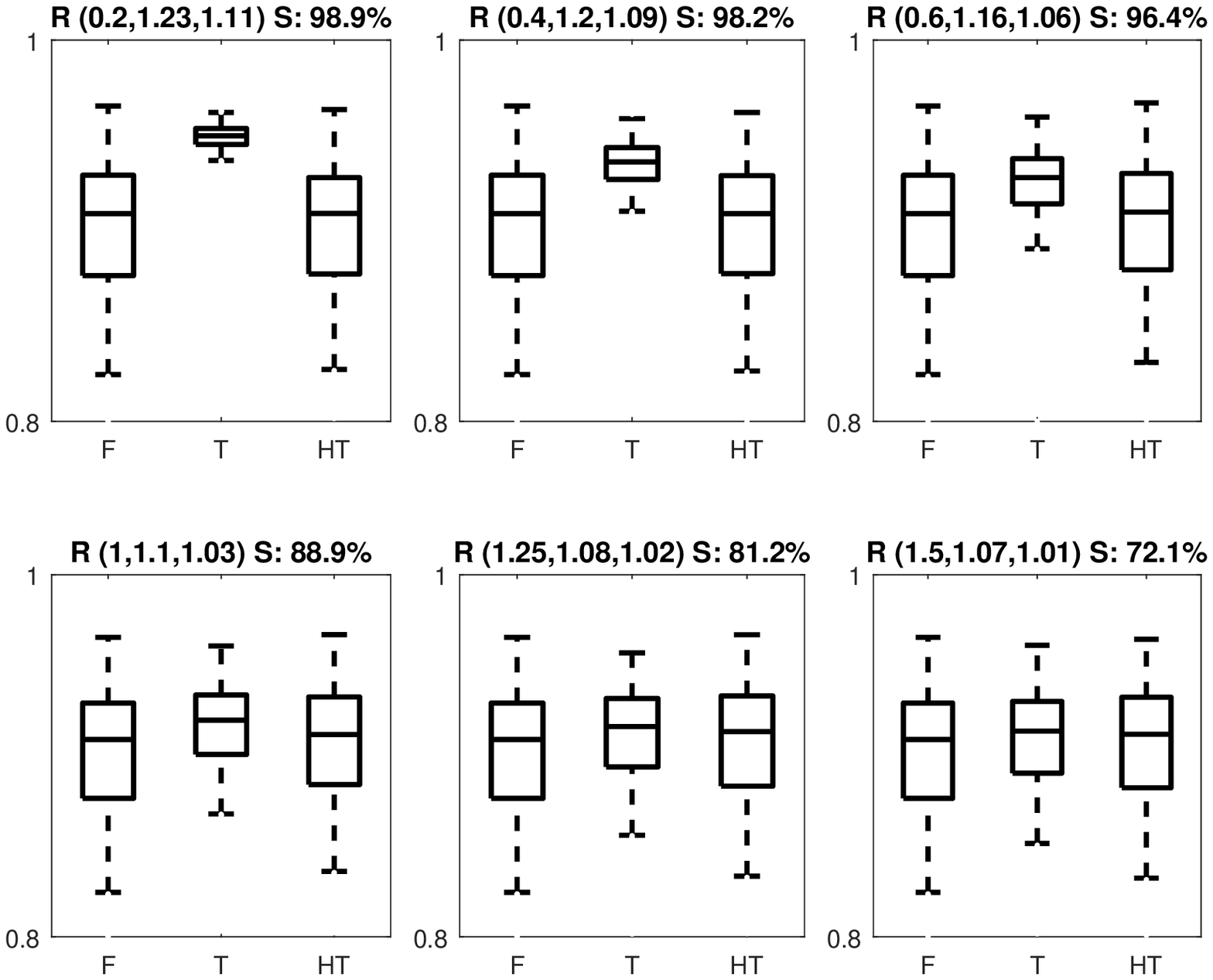} 
\caption{Boxplots of the 3D connectivity at $g(0.2)$ for approaches F, T and HT and for various ratios of taper range/effective range. Ratios in the subtitles are: (taper range/effective range, $\frac{MSE^s(x,C_1)}{MSE^s(x,C_0)}$, $\frac{MSE^s(x,C_0,C_1)}{MSE^s(x,C_0)}$). Experimental sparsity is indicated. Desired covariance exponential with effective range $16\frac{2}{3}$, spherical taper. Forty conditional realizations $\times$ 40 independent samples.}
\label{fig:taper_exp_spher_3d_allard_50}
\end{figure}

We now explore the connectivity in 3D. It is known (see e.g. \citep{renard2013}) that in 3D percolation holds at lower proportions than in 2D. We thus generate 3D realizations on a $50 \times 50 \times 50$ grid of truncated Gaussian random fields with $p=0.2$. The desired covariance,  $C_0$, is exponential with effective range $16\frac{2}{3}$. The taper is the spherical covariance. Forty samples of size 100, taken from as many different reference fields, are used to provide each 40 realizations.  Figure \ref{fig:taper_exp_spher_3d_allard_50} shows the distributions of the connectivity measure $g(0.2)$ for conditional simulations obtained from methods F, T and HT. Results indicate a clear discrepancy between connectivity distributions obtained by F and T methods. On the contrary, distributions obtained with HT are similar to those with F, even for a strong tapering inducing over $98\%$ sparsity.

\section{Discussion and Conclusion}

We explored theoretically and on simulations a Half-Tapering approach in which non conditional simulations are done with the desired covariance function $C_0$ using efficient methods such as turning bands or circulant embedding methods. The conditioning step is then performed  using a tapered covariance function $C_1 = C_0.C_T$. The HT method, already suggested in \cite{stein2013} enables fast conditional simulation free of neighborhood discontinuities, even with very large datasets.

The $MSE^s$ criterion enables to evaluate the quality of the approximation obtained with the HT approach compared to the F approach. We have extended to the simulation case the asymptotic results that were established in \cite{furrer2006} and extended in \cite{stein2013} for Best Linear Unbiased Prediction, thereby offering solid theoretical grounds to this approach. It was proven that, contrary to the T approach, the HT method is asymptotically equivalent to the F method. Expression to approximate the sparsity of the covariance matrix corresponding to a given taper range were derived. Hence, both the quality of the approximation and the corresponding sparsity can be assessed beforehand to find the best compromise for the taper range to use in the HT approach.  

The proposed HT approach keeps the efficiency and low computational complexity of most efficient unconditional simulation algorithms and benefit of the recent advances on tapering functions for the post-conditioning by kriging. As an example, the computational complexity of the turning bands unconditional simulation is only $O(N)$ where $N$ is the number of points to simulate. For the post-conditioning, the memory requirement is inversely proportional to the matrix sparsity. Moreover, best implementations of sparse Cholesky decomposition used to solve the kriging system have computational complexity proportional to the number of non-zero elements instead of $O(n^3)$ for full matrices, $n$ being the number of conditioning data (see \cite{davis2006_sparse}, \cite{chen2008_sparse}).
 
The impact of approximating the desired covariance by the full tapering (T) and our proposed half-tapering (HT) approach has been assessed in both 1D, 2D and 3D cases with four different responses functions. In all cases considered, HT appeared more similar to F than T for both the response distributions and the $MSE^s$ ratio. 

The examples show that for the same similarity with F, simulations with HT allow shorter taper range and larger sparsity than T. In mining applications, the deposit is usually 3D. It is common that a few ten thousand data are available. Moreover, the effective range of the desired covariance can be quite smaller than the size of the field. These are all favorable conditions for HT. It enables to post-condition the simulation quickly at once, without introducing undesirable discontinuities like in the traditional approach based on local neighborhoods.

All the examples indicated a direct relation between the $MSE^s$ and the similarity of T or HT responses with F responses. The $MSE^s$ ratios for T were substantially larger than for HT. Increasing the taper range normally reduces both $MSE^s$ ratios. One noticeable exception is observed in Fig. \ref{fig:mse} where the combination cubic covariance-penta taper shows a slight MSE ratio increase with an increase of the taper range. However, this strange behavior was not observed in the cases of Wendland$_1$ and cubic tapers. Hence, our results support the use of taper functions having the same differentiability at origin and spectrum decay rate as the desired covariance. 

For all forward models considered, a $MSE^s$ ratio below 1.1 for HT was shown sufficient to provide similar responses to F on the non linear responses. On the contrary, for T, such a threshold on $MSE^s$ could not be established (see Figs \ref{fig:taper_exp_spher_max}, \ref{fig:taper_exp_spher} and \ref{fig:taper_exp_spher_3d_allard_50}).

This work could be generalized in several directions that we now briefly mention. In the few cases where tapering alone does not suffice to bring enough sparsity, a possible extension to our approach would be to complete the tapering by using also fixed rank kriging, similarly as done in \cite{sang2012}. The hybrid approach applied for the post-conditioning by kriging would enable to extend the computational advantages of HT to most situations.
 
Also, in this work, we have only considered the case of evenly spaced conditioning points. \cite{bolin2013comparison} showed that tapering can perform poorly for kriging when data points are very irregularly located. The reason is that performances of the tapering approach is not only a function of the ratio of ranges of the taper and the desired covariance but also of the  average spacing of observations. The optimal taper range achieves a balance between computational cost and accuracy. In regions with sparse observations the taper range could be increased to obtain lower $MSE$ ratios without penalizing significantly the sparsity. In regions with high density of data, the computation burden could be reduced. In \cite{bolin2016}, a spatially adaptive covariance tapering strategy where the taper range is allowed to vary in space is proposed to improve the performance of tapering. This strategy could also be implemented for conditional simulations when the conditioning data are very irregularly located. As an example, with 3D borehole data where sampling density is larger along borehole direction it might be required to use anisotropic tapers with shorter range parallel to the borehole axis.

An interesting alternative \citep{porcu2016wendland} to the tapering of the desired covariance would be to use directly the taper function as the covariance in the post-conditioning by kriging. Properties of this approach remain to be established. However, it seems quite clear that the differentiability and spectrum decay rate of the taper and the tapered covariance must be the same. In a few simulation tests, not shown here for conciseness, we obtained very similar results  with the tapered covariance and the taper function alone. 

Using a tapering approach when $C_0$ is a Cauchy or a Gaussian covariance function remains an open problem. For these covariance functions, the spectral density behave either as $s^\alpha \exp\{-s\}$ or as $\exp\{-s^2\}$ as $s \to \infty$. The tail condition, which is a sufficient condition seems to be overly demanding. New theoretical results for finding necessary conditions for tapering, or at least less demanding sufficient conditions must be established.

Multivariate extension of this work is relatively straightforward. It suffices to do the covariance tapering with a positive semi-definite multivariate model defined on compact support. The linear model of coregionalization \citep{chiles2012,wackernagel2003} is an easy way to obtain such multivariate taper. Moreover, \cite{porcu2013radial} defined a class of multivariate radial covariances based on constructions involving Askey and Buhmann functions. They provided sufficient conditions for the positive definiteness of the multivariate models in this class.

\section*{Appendix: Proof of Proposition  \ref{prop:prop3}}

We first establish two lemmas:

\begin{Lemma}[Adapted from \citep{lu2002}]
Consider the symmetric block matrix 
$$\left( \begin{array}{cc} \bK_0^{-1} & \bD_T \\
\bD_T & \bK_1\end{array}\right),$$
where $\bD_T$ is diagonal and $\bK_0$ and $\bK_1$ are symmetric non singular matrices such that 
$\bK_1 - \bD_T \bK_0 \bD_T$ and $\bK_0^{-1} - \bD_T \bK_1 \bD_T$ are non singular. Then, 
$$(\bK_0^{-1} - \bD_T \bK_1 \bD_T)^{-1} = \bK_0 + \bK_0 \bD_T (\bK_1 - \bD_T \bK_0 \bD_T)^{-1} \bD_T \bK_0.$$
\end{Lemma}
\noindent{\bf Proof} \quad This lemma is a direct consequence of Theorem 2 in \cite{lu2002}. \finpreuve

\begin{Lemma}
Let $\bx_1,\dots, \bx_n$ be $n$ sample points in finite domain $D$ and let $C(\bh)$ be a covariance function on $D$ with $C(\bzero)=1$. Let $Z$ be a zero mean Gaussian random field with covariance function $C(\bh)$ on $D$.  Let further $\bK$ be the $n \times n$ matrix with elements
$[\bK]_{ij} = C(\bx_i-\bx_j)$, for $1 \leq i,j \leq n$ and $\bk_s$ be the $n$ vector with elements $[\bk_\bx]_i = C(\bx_i-\bx)$, for $\bx \in D$ and $1 \leq i \leq n$. Then,
the matrix 
$$\bK - \bk_\bx \bk'_\bx$$ is positive semi-definite for any $\bx \in D$. 
\end{Lemma}

\noindent {\bf Proof} 
\quad In order to show this, we need to show that for any vector $\blambda = (\lambda_1,\dots,\lambda_n)' \in \Re^n$, it holds that
\begin{equation}
Q = \sum_{i=1}^n \sum_{j=1}^n \lambda_i \lambda_j \left( [\bK]_{ij}  - [\bk_\bx]_{i} [\bk_\bx]_{j} \right)\geq 0.
\label{eq:lemma2}
\end{equation}
Let us denote $S = \sum_{i=1}^n  \lambda_i Z(\bx_i)$. Then, since $\hbox{Var}\{S\} = \sum_{i=1}^n \sum_{j=1}^n \lambda_i \lambda_j  [\bK]_{ij}$ and since
$\hbox{Cov}\{S,Z(\bx)\} = \sum_{i=1}^n \lambda_i [\bk_\bx]_i$, Eq. (\ref{eq:lemma2}) is  equivalent to 
$$Q = \hbox{Var}\{S\} -\hbox{Cov}\{S,Z(\bx)\}^2.$$
Using $\hbox{Cov}\{S,Z(\bx)\}^2 \leq  \hbox{Var}\{S\}  \hbox{Var}\{Z(\bx)\}$ and $\hbox{Var}\{Z(\bx)\} = 1$, we thus get very easily that
$$Q \geq \hbox{Var}\{S\} -\hbox{Var}\{S\}  \hbox{Var}\{Z(\bx_0)\} = 0,$$
which finishes the proof. \finpreuve

\bigskip

We are now ready to provide the proof of Proposition \ref{prop:prop3}. We must show  that $\sigma^2_{k,C_1}(\bx) \geq \sigma^2_{k,C_0}(\bx)$ for all $\bx \in D$. As usual,  we drop the dependency on $\bx$ for sake of conciseness. Since $\sigma^2_{k,C_1} = \sigma^2_0 - \bk'_1 \bK_1^{-1} \bk_1$ and
$\sigma^2_{k,C_0} = \sigma^2_0 - \bk'_0 \bK_0^{-1} \bk_0$, we need to prove that:
\begin{equation}
\bk'_0 \bK_0^{-1} \bk_0 - \bk'_1 \bK_1^{-1} \bk_1 \geq 0.
\label{eq:proof1}
\end{equation}
Since $\bk_1 = \bk_0 \bk_T$ and $\bK_1 = \bK_0 \odot \bK_T$, Eq. (\ref{eq:proof1}) is equivalent to 
\begin{equation}
\sum_{i=1}^n \sum_{j=1}^n [\bk_0]_i \left( [\bK_0^{-1}]_{ij} - [\bk_T]_i \, [\{\bK_0 \odot \bK_T\}^{-1}]_{ij}  \, [\bk_T]_j\right) [\bk_0]_j \geq 0.
\end{equation}
To show that this expression is always nonnegative, we will show that the matrix $\bM$ with elements
$[\bM]_{ij} = [\bK_0^{-1}]_{ij} - [\bk_T]_i \, [\{\bK_0 \odot \bK_T\}^{-1}]_{ij} \, [\bk_T]_j$, for $1 \leq i,j \leq n$ is  positive definite (p.d.) except for the trivial case $\bK_0=\bK_1, \bk_0=\bk_1$, corresponding to a taper with infinite range, where $\bM=\mathbf{0}$ and Eq. \ref{eq:proof1} equals zero. Introducing the diagonal matrix $\bD_T = \hbox{diag}(\bk_T)$, this matrix can also be written
\begin{equation}
\bM=\bK_0^{-1} - \bD_T\, \{\bK_0 \odot \bK_T\}^{-1}  \bD_T.
\end{equation}
Since $\bM$ is invertible, it is p.d. if and only if $\bM^{-1}$ is  p.d. Using Lemma 1, its inverse is 
\begin{equation}
\bM^{-1} = \bK_0 +  \bK_0 \bD_T\, \{\bK_0 \odot \bK_T - \bD_T \bK_0 \bD_T\}^{-1}  \bD_T  \bK_0.
\label{eq:proof2}
\end{equation}
Using Lemma 2, one has that $\bK_T - \bk_T \bk_T'$ is p.d. Hence, using Schur's product theorem, 
$\bK_0 \odot (\bK_T - \bk_T \bk_T') = \bK_0 \odot \bK_T - \bK_0 \odot \bk_T \bk_T'= 
\bK_0 \odot \bK_T -  \bD_T \bK_0 \bD_T$ is p.d. and so is its inverse. As sums and products of p.d. matrices are p.d., we can conclude that $\bM^{-1}$ in Eq. (\ref{eq:proof2}) is also p.d., which completes the proof. \finpreuve

\subsection*{acknowledgements}
We are indebted to one anonymous reviewer for his attentive and detailed review and for his numerous constructive comments. We thank Pr. Emilio Porcu from Universidad T\'ecnica Federico Santa Mar\'\i a in Valparaiso (Chile) for fruitful discussions and for providing us working material on generalized Wendland covariance functions and their use under fixed domain asymptotics. This research was financed in part by National Science Research Council of Canada (grant RGPIN105603-05)


\end{document}